\documentclass[aps, prb, twocolumn, floatfix, letterpaper, nofootinbib ]{revtex4-2}

\usepackage[a-1b]{pdfx}

\usepackage{graphicx}
\usepackage{dcolumn}
\usepackage{bm}
\usepackage{multirow}

\usepackage[normalem]{ulem}

\usepackage{amsmath}
\usepackage{amsthm}
\usepackage{amstext}
\usepackage{amssymb}
\usepackage{mathrsfs}
\usepackage{amsfonts}
\usepackage{amsbsy} 

\usepackage[all,matrix,cmtip]{xy}

\usepackage{csquotes}
\MakeOuterQuote{"}

\usepackage{color}

\definecolor{red}{rgb}{1,0,0}
\definecolor{blue}{rgb}{0,0,1}
\definecolor{dblue}{rgb}{0,0,0.4}
\definecolor{green}{rgb}{0,1,0}
\definecolor{dgreen}{rgb}{0,0.4,0}
\definecolor{black}{rgb}{0,0,0}
\definecolor{white}{rgb}{1,1,1}

\definecolor{brn}{rgb}{.8,.4,.0}
\definecolor{redo}{rgb}{1,.5,.0}
\definecolor{ddgrn}{rgb}{0,0.4,0}
\definecolor{dgrn}{rgb}{0,0.55,0}
\definecolor{dbl}{rgb}{0,0,0.5}
\definecolor{grey}{rgb}{0.5,0.5,0.5}

\usepackage[bbgreekl]{mathbbol}

\renewcommand{\v}[1]{\boldsymbol{#1}} 
 
\newcommand{\ii}{\hspace{1pt}\mathrm{i}\hspace{1pt}}

\newcommand{\dd}{\hspace{1pt}\mathrm{d}}
 
\renewcommand{\>}{\rangle} 

\newcommand{\Rf}[1]{Ref.~\onlinecite{#1}}

\newcommand{\ie}{{\it i.e.~}}

\newcommand{\bpm}{\begin{pmatrix}}
\newcommand{\epm}{\end{pmatrix}}
\newcommand{\bmm}{\begin{matrix}}
\newcommand{\emm}{\end{matrix}}

\newcommand{\cP}{ {\cal P} }

\usepackage{euscript}

\newcommand\sT         {\mathsf{T}}

\newcommand{\al}{\alpha} 
\newcommand{\bt}{\beta} 
\newcommand{\del}{\delta} 
\newcommand{\Del}{\Delta} 
\newcommand{\eps}{\epsilon} 
 
\newcommand{\ga}{\gamma}

\newcommand{\si}{\sigma}

\usepackage{longtable}

\usepackage{hyperref}
\hypersetup{
    colorlinks = true,
    allcolors = blue,
}

\begin{document}

\title{Non-Abelian Fibonacci quantum Hall states in tetralayer rhombohedral graphene}

\author{Abigail Timmel}
\affiliation{Department of Physics, Massachusetts Institute of Technology,
Cambridge, Massachusetts 02139, USA}

\author{Xiao-Gang Wen}
\affiliation{Department of Physics, Massachusetts Institute of Technology,
Cambridge, Massachusetts 2139, USA}

\begin{abstract} 

In 1991, it was proposed that fourfold-degenerate Landau levels formed by a
single species of electrons could host a non-Abelian fractional quantum Hall
(FQH) state with Fibonacci anyons at filling fraction $\nu = \frac{2}{3}$. In
this work, we investigate how such degenerate Landau levels can be realized in
rhombohedral-stacked tetralayer graphene. We identify the following key
conditions which may stabilize the Fibonacci state: (1) A magnetic field of around
20 Tesla is required if surface and interior carbons have the same energy
level.  If substrate hybridization raises the surface carbon energy level by
$\Delta_2 = 30$\,meV relative to interior carbon, the required field will have
a larger range:  15 -- 20 Tesla. For $\Delta_2 = 45$\,meV, the range reaches a
maximum: 7 -- 20 Tesla.  (2) The displacement field must be tuned to achieve
Landau level degeneracy.  $\nu = \frac{3}{5}$ Fibonacci FQH states may also
be realized in pentalayer rhombohedral graphene with a magnetic field of 12 Tesla,
and $\nu = \frac12$ states with Ising anyons may occur in trilayer graphene for 
magnetic fields of 12 -- 20 Tesla at $\Delta_2 = 0$ or 5 -- 20 Tesla at 
$\Delta_2 = 45$\,meV.  We also study a simple interaction model to 
explore spin/valley polarization effects, and we see that the Fibonacci state
may occur at $\nu = 2/3 + $ integer filling fractions, where the integer is
0 and 4 for sufficiently weak interaction, or can shift to 2 and 5 under 
a stronger interaction. The case $\Delta_2 = 45$\,meV also produces states at 
negative filling fraction, e.g. $-\frac23$, $-4\frac23$.  Here $\nu$ is defined 
with respect to the Hall conductance, $\sigma_{xy} = \nu \frac{e^2}{h}$.

\end{abstract}

\maketitle


\section{Introduction}

Particle exchange statistics come with distinct features that 
have profound implications in our world.  In two
spatial dimensions, representations of the braid group
\cite{LM7701,W8257,W8413} can go beyond bosonic and fermionic,
offereing more mathematical possibilities for particle exchange statistics. Abelian
statistics are described by 1-dimensional representations of braid group, and
\Rf{H8483,ASW8422} showed that these can be realized in electron systems as
quasiparticle excitations.  Even for non-Abelian statistics
\cite{W8413,GMS8503} described by higher-dimensional representations of braid
group, \Rf{W9102,MR9162} showed that they can be realized in electron or qubit
systems as quasiparticle excitations.  Such non-Abelian states of electrons or
qubits may open pathways for topological quantum computation.  Fibonacci
statistics in particular are the simplest non-Abelian exchange supporting
universal quantum computation \cite{FKL0331,FLZ0205}.  Discovering non-Abelian
fractional quantum Hall (FQH) states with  Fibonacci statistics will be an important
milestone in experimental realizations of exotic quantum states.

The filling fraction $\nu=1/3$ Laughlin state \cite{L8395}
\begin{align}
 \Psi(\{z_i\})=
\chi_1(\{z_i\})
\chi_1(\{z_i\})
\chi_1(\{z_i\}) , \
z_i=x_i+\ii y_i
\end{align}
describes a gapped quantum Hall (QH) state that contains anyons with
fractional charges \cite{TSG8259,L8395,dRH9762}, Abelian fractional statistics
\cite{LM7701,W8257,H8483,ASW8422,W8413}, and topological order
\cite{W8987,W9039,WN9077}.  Here $\chi_n(\{z_i\})$ is the fermion wave function
of $n$-filled Landau levels.  \Rf{J8979} proposed that a generalized wave
function $\Psi=\chi_{n_1}\chi_{n_2}\chi_{n_3}$ can be gapped for small 
$n_i$'s.

Two independent proposals to realize non-Abelian statistics in quantum Hall
systems emerged around the same time.  \Rf{W9102,W9927} pointed out that the
FQH states described by wave functions
\begin{align}
\label{nabLL}
 \Psi_{\nu=\frac{n}{m}}(\{z_i\})&=[\chi_n(\{z_i\})]^m, 
\nonumber\\
\text { or }
 \Psi_{\nu=\frac{n}{m+n}}(\{z_i\})&=\chi_1(\{z_i\})[\chi_n(\{z_i\})]^m 
\end{align}
are described by low energy effective field theory of level-$n$ SU$(m)$
Chern-Simons theory.  Thus they support topological excitations with
non-Abelian statistics of $SU(m)_n$-type.  (To be more precise, the low energy
effective theory is spin-$SU(m)_n$ Chern-Simons theory, since the
$SU(m)$-singlet can be fermions -- the electrons.) The above two states both
have non-trivial edge states described by $U(1)\times SU(n)_m$ Kac-Moody
current algebra \cite{BW9215}. 

Note that, according to the power of $\bar z$'s, the $\nu = 2/3$ state
described by $[\chi_2({z_i})]^3$ and the $\nu = 3/5$ state described by
$\chi_1[\chi_3({z_i})]^2$ occupy only the first four and first five Landau
levels, respectively. These wavefunctions represent exact ground states for
electrons interacting via $\delta(z_i - z_j)$ repulsion, provided
that the first four or five Landau levels are degenerate.  Consequently,
$[\chi_2({z_i})]^3$ and $\chi_1[\chi_3({z_i})]^2$ are primary stable
fractional quantum Hall (FQH) states under those conditions \cite{W9102, W9927,
TO220409684}. Remarkably, both states support Fibonacci anyons.

Using a conformal field theory (CFT) approach, \Rf{MR9162} proposed that the
Pfaffian wave function
\begin{align}
\Psi_{\nu=1/2} =
\mathrm{Pf}\left[ \frac{1}{z_i - z_j} \right]
e^{-\frac{1}{4} \sum |z_i|^2}
\prod (z_i - z_j)^2
\end{align}
describes a gapped FQH state. Since the Pfaffian wave function corresponds to
the correlation function of a primary field in the Ising CFT, it was
conjectured that the associated FQH state supports excitations with non-Abelian
statistics of the Ising type (or $SU(2)_2$ type) \cite{MR9162}.

The edge states of the Pfaffian state were studied numerically in \Rf{W9355}
and were found to be described by a $c=1$ chiral boson CFT and a $c=1/2$ chiral
Majorana fermion CFT, supporting the proposal that the Pfaffian state is indeed
non-Abelian.  Subsequent confirmations of its non-Abelian property include:
computations of quantum dimensions \cite{NW9629}, derivations of low-energy
effective theories such as the $SO(5)_1$ Chern-Simons theory \cite{W9927} and
the orbifold $U(1)^2$ Chern-Simons theory \cite{FHZ0191}, and a plasma analogy
calculation \cite{BN10085194}.  Furthermore, using the $\mathbb{Z}_3$
parafermion CFT, one can construct wave functions that support Fibonacci anyons
\cite{RR9984}.

In \Rf{FB200600238}, a $\nu=3/7$ FQH state, $\cP[(\chi_3^*)^2\chi_1^3]$, is
shown to be gapped and is proposed to have  $SU(2)_3$-type non-Abelian statistics,
where $\cP$ is the projection to the lowest Landau level.  However, the
unprojected wave function is quite complicated containing both $\chi_n$ and
$\chi_m^*$, and the projection $\cP$ is a large deformation of the
wave function, which may destroy the $SU(2)_3$ non-Abelian topological order
while maintaining the gap.  More direct numerical calculations, such as the ground
state degeneracy on a torus and/or the entanglement edge spectrum on disk, are needed
to confirm the projected state $\cP[(\chi_3^*)^2\chi_1^3]$ to be non-Abelian.
Similarly, \Rf{BB230202478} proposed that an even more complicated wave
function, a $\nu=4/11$ FQH state $\cP[\chi_4^*(\chi_2^*)^2\chi_1^4]$, has
$SU(2)_2$-type non-Abelian statistics.  Again more direct numerical
calculations are needed to confirm the projected state
$\cP[\chi_4^*(\chi_2^*)^2\chi_1^4]$ to be non-Abelian.  In contrast, the wave
functions Eq.~\eqref{nabLL} for degenerate Landau levels do not need to be
projected, and their non-Abelian character is more reliable.

Experimentally, the filling fraction $\nu=\frac12 +\mathrm{integer}$ FQH state
was first discovered in semiconductor samples \cite{WES8776,XWc0406724}.  The
quarter-electron charged quasi-particle was observed \cite{DM08020930} in such
state.  Both the Abelian (331) state and the non-Abelian Pfaffian and $SU(2)_2$
states have quarter-charged quasi-particles, and thus the observation of
quarter electron charge does not imply non-Abelian statistics.  However, the
observation of half-integer thermal Hall conductance suggests the $\nu=\frac12
+ $ integer FQH state to be a non-Abelian state \cite{BS171000492}.  The
filling fraction $\nu=\frac12 + \mathrm{integer}$ FQH state was also observed
in bilayer graphene \cite{KM13054761,ZY161107113,HZ210507058} and in wide
quantum well \cite{SS230900111}.  The measured daughter states
\cite{ZY161107113,HZ210507058,SS230900111} of filling fraction $\nu=\frac12 +
\mathrm{integer}$ FQH state provides a strong indication that the FQH state is
the non-Abelian (anti)-Pfaffian state \cite{LH08120381,ZW240612068}.

In this paper, we propose using 3,4,5-layer graphene with rhombohedral stacking
under a magnetic field to realize new non-Abelian FQH states of the form given
in Eq.~\eqref{nabLL}, corresponding to $SU(2)_2$, $SU(3)_2$, and $SU(2)_3$
topological orders. These states are expected to be the primary stable FQH
phases in such systems. Notably, rhombohedrally stacked multilayer graphene
($3$, $4$, and $5$ layers) is currently under active experimental investigation
\cite{YJ220212330,HJ230503151,LC230611042}.

We remark that the $SU(2)_2$ Ising non-Abelian state in trilayer graphene with
rhombohedral stacking was previously studied theoretically in \Rf{WJ160302153},
using a minimal model for the graphene band structure. The minimal model has
exact Landau level degeneracy. In contrast, the present work considers the
realistic band structure of graphene where the Landau level degeneracy is not
guaranteed. We aim to identify the experimental conditions that could favor or
suppress the emergence of $SU(3)_2$ and $SU(2)_3$ non-Abelian states
featuring Fibonacci anyons. Specifically, we examine the circumstances under
which real graphene can effectively simulate degenerate Landau levels of a
single species.  

We emphasize that degeneracies involving Landau levels from different
spin-valley quantum numbers are not suitable for realizing the $SU(m)_n$
non-Abelian states described by Eq.~\eqref{nabLL}.  We need degenerate Landau
levels from a single species of electrons.

\section{Minimal model for multi-layer rhombohedral graphene}


Using a minimal model for band structure, \Rf{MM07114333} found that
$n$-degenerate Landau levels can be realized by $n$-layer of graphene with
rhombohedral stacking under a magnetic field.  \Rf{MM07114333} also pointed out
that longer range hopping beyond the  minimal model can destoy the Landau level
degeneracy.  Rhombohedral stacking is a configuration of graphene layers where
the $A$ sublattice of one layer lies directly under the $B$ sublattice of the
next layer.  The displacements continue in the same direction moving up the
stack, which is in contrast to the alternating case of Bernal graphene.  Three
displacements by the sublattice separation vector correspond to a lattice
vector of graphene, so rhombohedral stacking follows an $ABC$ pattern where
every third layer has zero relative displacement.

A single layer of graphene has the low energy effective Hamiltonian
\begin{align}
 H = \hbar v_F \; {\bf k} \cdot \boldsymbol{\sigma} \oplus c.c.
\end{align}
where $v_F$ is the Fermi velocity, ${\bf k} = (k_x,k_y)$, and $\boldsymbol{\sigma}$ is the vector of Pauli matrices.  The direct sum accounts for two inequivalent valleys at the $K$ and $K'$ points which are time-reversed partners of each other.

We introduce the magnetic field by promoting the crystal momenta $k_j$ to
operators $i\partial_j$ and coupling to a vector potential $A_j$.  Let
\begin{align*}
	& \pi_j = \ii \partial_j + \frac{e}{\hbar c}A_j \\
	& \hat a = \frac{(\pi_x + \ii \pi_y)}{\sqrt{2eB/\hbar c}}, \quad [a,a^\dagger] = 1
\end{align*}
using cgs units.  Here $\hat a$ and $\hat a^\dagger$ are the usual raising and lowering operators for Landau levels.  The Hamiltonian can be rewritten in terms of these as

\begin{align}
 H =  \kappa \left( \begin{pmatrix}
	0 & \hat a^\dagger \\
	\hat a & 0
\end{pmatrix} \oplus 
\begin{pmatrix}
	0 & \hat a \\
	\hat a^\dagger & 0 
\end{pmatrix} \right)
\end{align}
where $\kappa = v_F \sqrt{2eB\hbar /c}$.  We use a basis of Landau levels, i.e.
eigenstates of the number operator $a^\dagger a |n\> = n|n\>$, to express
states in this formalism.  We can immediately see that a state $\psi^\sT =
(|0\>, 0)$ is annihilated in the first valley, and another state $\psi'^\sT =
(0, |0\>)$ is annihilated in the second valley.  These two states have zero
energy, and they are sublattice and valley polarized.  This is consistent with
the magnetic field breaking time-reversal symmetry but preserving inversion
symmetry.  The sublattice separation and valley separation vectors are
$90^\circ$ rotated from each other, so sublattice/valley polarization provides
a picture of cyclotron motion about the graphene unit cell center of mass.

Now we stack the layers rhombohedrally, and include only intra-layer and
interlayer nearest neighbor hopping (\ie consider the minimal-model for
graphene bands).  The Hamiltonian in the $K$ valley (suppressing the $K'$
valley for convenience) is of the form
\begin{align}
H = \kappa \begin{pmatrix}
	0 & D^\dagger \\
	D & 0 
\end{pmatrix}, \quad
\label{fredholm}
\end{align}
This can be deduced from the bipartite nature of the lattice: $A$ sublattices
only couple to $B$ sublattices and vice versa, so a basis
$\{A_1,\ldots,A_n\,B_1,\ldots,B_n\}$ results in the form Eq.~\eqref{fredholm}.  The
form of $D$ for the ideal rhombohedral stacking model has
\begin{align}
D = \begin{pmatrix}
	\hat a & \eta & 0  & \cdots \\
	0 & \hat a & \eta & \\
	\vdots & & \ddots & \ddots 
\end{pmatrix}
\end{align}
where $\eta = \ga_1/(v_F\sqrt{2e\hbar B/c})$. Here $v_F = \frac32 \frac{\ga_0
a}{\hbar}$ is the velocity at the Fermi point, $a=1.42$\AA\ is the
carbon-carbon distance, $\ga_0 = 3.1$\,eV is the intra-layer nearest neighbor hopping
energy, and $\ga_1 = 0.3 - 0.4$\,eV is the interlayer nearest neighbor hopping
energy \cite{ZM08091898}.  From this, we can find the zero energy subspace
$\psi_0$, which are states annihilated by $D$.  
\begin{align}
	\label{wfn}
	 \psi_0 = \mathrm{span}_{\ell=0}^{n-1}\left\{\sum_{k=0}^\ell \tfrac{(-1)^{k}\sqrt{\ell!}}{\eta^k\sqrt{(\ell-k)!}}|\ell-k\>|A_{k+1}\> \right\}
\end{align}

As we saw in the one layer case, these states live only on the $A$ sublattice.
If we were to look at the $K'$ valley, we would find that those zero energy
states live only on the $B$ sublattice.  Furthermore, we can see that the $K$
valley favors the top layer in the large $\eta$ limit (magnetic field weak
compared to interlayer coupling).  Similarly, the $K'$ valley favors the bottom
layer.  For later convenience, we introduce an operator $W$ that maps the above
wave function for $\eta = \infty$ to those for finite $\eta$:
\begin{align}
	W |\ell\> = \sum_{k=0}^\ell \tfrac{(-1)^{k}\sqrt{\ell!}}{\eta^k\sqrt{(\ell-k)!}}|\ell-k\>|A_{k+1}\>
\end{align}

The zero-energy subspace has dimension $n$, and its basis consists of Landau
levels $1 \ldots n$.
The asymmetry in number of zero-energy states living on the $A$ vs. $B$
sublattices is a protected quantity known as the index of $D$, given by $\dim
\ker D - \dim \ker D^\dagger$.  This index and the resulting Landau level
degeneracy is robust to perturbations that preserve the form of $H$ in
Eq.~\eqref{fredholm}, which is enforced by a symmetry between positive and negative
energies.  This symmetry is implemented by the transformation $|i_A\>,|i_B\>
\rightarrow |i_A\>,-|i_B\> $, so any hopping between $A,B$ sublattices will
preserve the degeneracy.  Examples of such $AB$-hopping that appear in more
realistic models of rhombohedral-stacked graphene are a $\gamma_2$ term,
coupling $A_i$ to $B_{i+2}$, and a $\gamma_3$ term, coupling $A_i$ to
$B_{i+1}$~\cite{GS221102492}.  
The $n$-fold degeneracy at zero energy will remain unchanged after adding not
so large $AB$-hoppings.  Realistic models of stacked graphene also include
$A$-$A$ and $B$-$B$ couplings which break this degeneracy, notably a $\gamma_4$
term which which couples $A_i$ to $A_{i+1}$ and $B_i$ to $B_{i+1}$.  We will
explore the effect of this term and how to counter the degeneracy splitting 
caused by those terms in the next section.

The existence of this zero energy subspace is not enough to realized the $SU(m)_n$
non-Abelian states in Eq.~\eqref{nabLL}.  We also require the degenerate Landau
levels to have a so-called single species property.  To define the single species property, we
consider a wave function $\Psi(\al,\v i)$ in the degenerate subspace and project
it to the $m$th Landau level, $\mathcal{P}_m \Psi(\al,\v i) = \Psi_m(\al,\v i)$,
where $\al$ labels graphene layers and $\v i$ labels the sites in a graphene
layer.  We can treat $\al$ and $\v i$ as independent degrees of freedom and
construct reduced density matrix by tracing out the $\v i$ degrees of freedom:
\begin{align}
 \rho^m_{\al\bt} =
\sum_{\v i} \Psi_m^*(\al,\v i) \Psi_m(\bt,\v i) .
\end{align}
The degenerate Landau levels labeled by $m_1,m_2 = 1,\cdots,n$
have the exact single species property
if they satisfy
\begin{align}
 \rho^{m_1} = \rho^{m_2} = \rho^{m_1} \rho^{m_2}, \ \ \
\forall \ m_1,m_2 \in [1,n]. 
\label{singlespecies}
\end{align}

This condition can fail in two ways: either the $m$th Landau level wave function is smeared 
across multiple layers, or the projected density matrix does not have trace 1 
(the original wave function had weight in other Landau levels).
In the minimal model, when $\eta \rightarrow \infty$, the states in the
degenerate Landau levels are localized entirely on the top layers for the
$K$-valley (and on the bottom layers for the $K'$-valley). In this limit, the
reduced density matrices $\rho^m_{\alpha\beta}$ satisfy the single-species
condition exactly. For large but finite $\eta$, this condition is only
approximately satisfied. However, computing the reduced density matrix in this
regime is challenging. Therefore, in this paper, we use the weight of
the wave functions in any of the top-layer first $N$ Landau levels as a proxy for how
well the single-species condition is fulfilled. The condition is considered
perfectly satisfied when this weight is one for each of the $N$ zero energy wavefunctions.

\section{Realistic model for multi-layer rhombohedral graphene}
\label{noninteracting}

Having identified the important issues using the minimal model for rhombohedral
graphene, we now turn to a more realistic model. This model incorporates up to
two-layer hopping with the smallest star of lateral translations and is
widely adopted for describing multilayer graphene systems.  We use the
numerical values from \cite{GS221102492}.  The hopping Hamiltonian has the
following form
\begin{align}
H = \begin{pmatrix} A & D^\dagger \\ D & B \end{pmatrix} 
\end{align}
where the basis is again $(A_1,\ldots,A_n,B_1,\ldots,B_n)$. 
For trilayer,
\begingroup
\allowdisplaybreaks
\begin{align}
 D &= \begin{pmatrix}
	v_0 \hat a &  \gamma_1 & 0 \\
        v_3 \hat a^\dagger & v_0 \hat a & \gamma_1  \\
        \frac{\gamma_2}{2} & v_3 \hat a^\dagger & v_0 \hat a \\
\end{pmatrix} 
\nonumber\\
 A &= \begin{pmatrix}
	0 & v_4 \hat a^\dagger & 0 \\
        v_4 \hat a & -\frac{u}{2}-\Delta_2+\delta & v_4 \hat a^\dagger \\
        0 & v_4 \hat a & -u+\delta \\
\end{pmatrix} 
\nonumber\\
B &= \begin{pmatrix}
	\delta & v_4 \hat a^\dagger & 0 \\
        v_4 \hat a & -\frac{u}{2}-\Delta_2+\delta & v_4 \hat a^\dagger \\
        0 & v_4 \hat a & -u \\
\end{pmatrix} 
\end{align}
For tetralayer,
\begin{align}
D &= \begin{pmatrix}
	v_0 \hat a &  \gamma_1 & 0 & 0 \\
        v_3 \hat a^\dagger & v_0 \hat a & \gamma_1 & 0 \\
        \frac{\gamma_2}{2} & v_3 \hat a^\dagger & v_0 \hat a & \gamma_1 \\
        0 & \frac{\gamma_2}{2} & v_3\hat a^\dagger & v_0 \hat a
\end{pmatrix}
\\
A &= \begin{pmatrix}
	0 & v_4 \hat a^\dagger & 0 & 0 \\
        v_4 \hat a & -\frac u3 -\Delta_2+\delta & v_4 \hat a^\dagger & 0 \\
        0 & v_4 \hat a & -\frac{2u}{3}-\Delta_2+\delta & v_4\hat a^\dagger \\
        0 & 0 & v_4 \hat a & -u+\delta
\end{pmatrix} 
\nonumber\\
B &= \begin{pmatrix}
	\delta & v_4 \hat a^\dagger & 0 & 0 \\
        v_4 \hat a & -\frac u3 -\Delta_2+\delta & v_4 \hat a^\dagger & 0 \\
        0 & v_4 \hat a & -\frac{2u}{3}-\Delta_2+\delta & v_4\hat a^\dagger \\
        0 & 0 & v_4 \hat a & -u
\end{pmatrix} 
\nonumber 
\end{align} 
For pentalayer,
\begin{align}
D &= \begin{pmatrix}
	v_0 \hat a &  \gamma_1 & 0 & 0 &0\\
        v_3 \hat a^\dagger & v_0 \hat a & \gamma_1 & 0&0 \\
        \frac{\gamma_2}{2} & v_3 \hat a^\dagger & v_0 \hat a & \gamma_1 &0\\
        0 & \frac{\gamma_2}{2} & v_3 \hat a^\dagger & v_0 \hat a & \gamma_1 \\
        0 &0 & \frac{\gamma_2}{2} & v_3\hat a^\dagger & v_0 \hat a
\end{pmatrix}
\\
A &= \footnotesize \begin{pmatrix}
	0 & v_4 \hat a^\dagger & 0 & 0 &0 \\
        v_4 \hat a & -\frac u4 -\Delta_2+\delta & v_4 \hat a^\dagger & 0 &0 \\
        0 & v_4 \hat a & -\frac{2u}{4}-\Delta_2+\delta & v_4\hat a^\dagger &0 \\
        0 &0 & v_4 \hat a & -\frac{3u}{4}-\Delta_2+\delta & v_4\hat a^\dagger \\
        0 &0 & 0 & v_4 \hat a & -u+\delta
\end{pmatrix} 
\nonumber\\
B &= \footnotesize \begin{pmatrix}
	\delta & v_4 \hat a^\dagger & 0 & 0 &0\\
        v_4 \hat a & -\frac u4 -\Delta_2+\delta & v_4 \hat a^\dagger & 0 &0\\
        0 & v_4 \hat a & -\frac{2u}{4}-\Delta_2+\delta & v_4\hat a^\dagger &0\\
        0& 0 & v_4 \hat a & -\frac{3u}{4}-\Delta_2+\delta & v_4\hat a^\dagger \\
        0& 0 & 0 & v_4 \hat a & -u
\end{pmatrix} 
\nonumber 
\end{align} 
\endgroup
The hopping Hamiltonian for $K'$-valley can be obtained from the above by
exchanging $\hat a \leftrightarrow \hat a^\dag$.

Here, for the spectrum under a magnetic field $B$, we use $v_i = \frac{3a}{2}
\sqrt{\frac{2eB}{\hbar c}} \gamma_i$, where $a = 0.142$ nm is the carbon-carbon
distance.  The hopping energies are $\gamma_0 = 3.1$\,eV, $ \gamma_1 =
0.38 $\,eV, $ \gamma_2 = -0.015 $\,eV, $ \gamma_3 = -0.29 $\,eV,
$\gamma_4 = -0.144 $\,eV.  

Three diagonal (on-site potential) terms $u,\ \Delta_2,\ \delta$ also appear in
this model.  They are defined such that the top layer, where near-zero
eigenstates have the most weight, is the zero of these potentials.  The
displacement field potential decreases from $0$ to $-u$ in equal increments
from the top layer to the bottom layer.  $-\Delta_2$ is the potential of the
inner layers relative to the surface layers.  Ref~\cite{GS221102492} finds that
the electrostatic contribution to this from the substrate can only be a few
meV. However, the hybridization with substrate may shift the energy level of
the surface layer relative to that of inner layers.  Thus $\Del_2$ is substrate
dependent.  Finally, $\delta$ is an electrostatic potential felt by
non-dangling bonds, \ie all sites except the top layer $A$ site and bottom
layer $B$ site.  $\delta$ is estimated at 10 meV.

The Landau level degeneracy is exact when $\ga_2 = \ga_4 = u = \Delta_2 =
\delta =0$.  $\ga_2$ and $\del$ are quite small but $\ga_4$ is large enough to
cause significant splitting in Landau level degeneracy.  Since the displacement
field $u$ is tunable in experiments, the goal will be to use $u$ to cancel the effect of
$\ga_4$.  It turns out that this cancellation is not perfect, but the 
splitting can be significantly reduced.  We also find that a positive $\Del_2=45$\,meV can cancel the effect of
$\ga_4$ almost perfectly, restoring the Landau level degeneracy.

To compute the Landau levels from $K$-valley, we utilize the lowering
operator $\hat a$ in a Landau level basis:
\begin{align}
 (\hat a)_{i,j} = \sqrt{i}\del_{j,i+1},\ \ \ \
i,j = 1,2,\cdots,\infty.
\end{align}
For numerical purposes, we need to truncate
the infinite dimensional Landau level basis to a finite
 dimensional subspace which can approximate the states near zero energy.  This 
can be done just by taking the lowest Landau levels up to some cutoff $N_{LL}$.
The lowering operator becomes
\begin{align}
 (\hat a)_{i,j} = \sqrt{i}\del_{j,i+1},\ \ \ \
i,j = 1,2,\cdots,N_{LL}.
\end{align}
Such a truncation of $\hat a$ produces spurious states near zero energy from 
$i,j$ near the cutoff $N_{LL}$, which we must eliminate.

To accomplish this, we add an imaginary
component to the Hamiltonian which is diagonal in the sublattice and layer
indices but increases smoothly to $\ii$ as the Landau level index goes to
$N_{LL}$.  More precisely, we add a regulator $R$,
\begin{align*}
	&H \rightarrow H + R \\
	&R_{ij} = \left\{ \begin{aligned}& \ii \delta_{ij} \frac{(j - n)^2}{(N_{LL}-n)^2}  \quad & j>n \\
		& 0 & \mathrm{otherwise}\end{aligned} \right.
\end{align*}
for some intermediate LL index $n$.  For the numerical calculations in this paper, we use $n=.7N_{LL}$.
High Landau level states will be penalized with a large imaginary component in their eigenvalues.
The low energy spectrum will then be naturally be concentrated in the lowest
Landau levels, which can be confirmed via the realness of their eigenvalues.


\subsection{Landau levels for tetralayer rhombohedral graphene}

\begin{figure}[t]
	a. \includegraphics[width=.9\columnwidth]{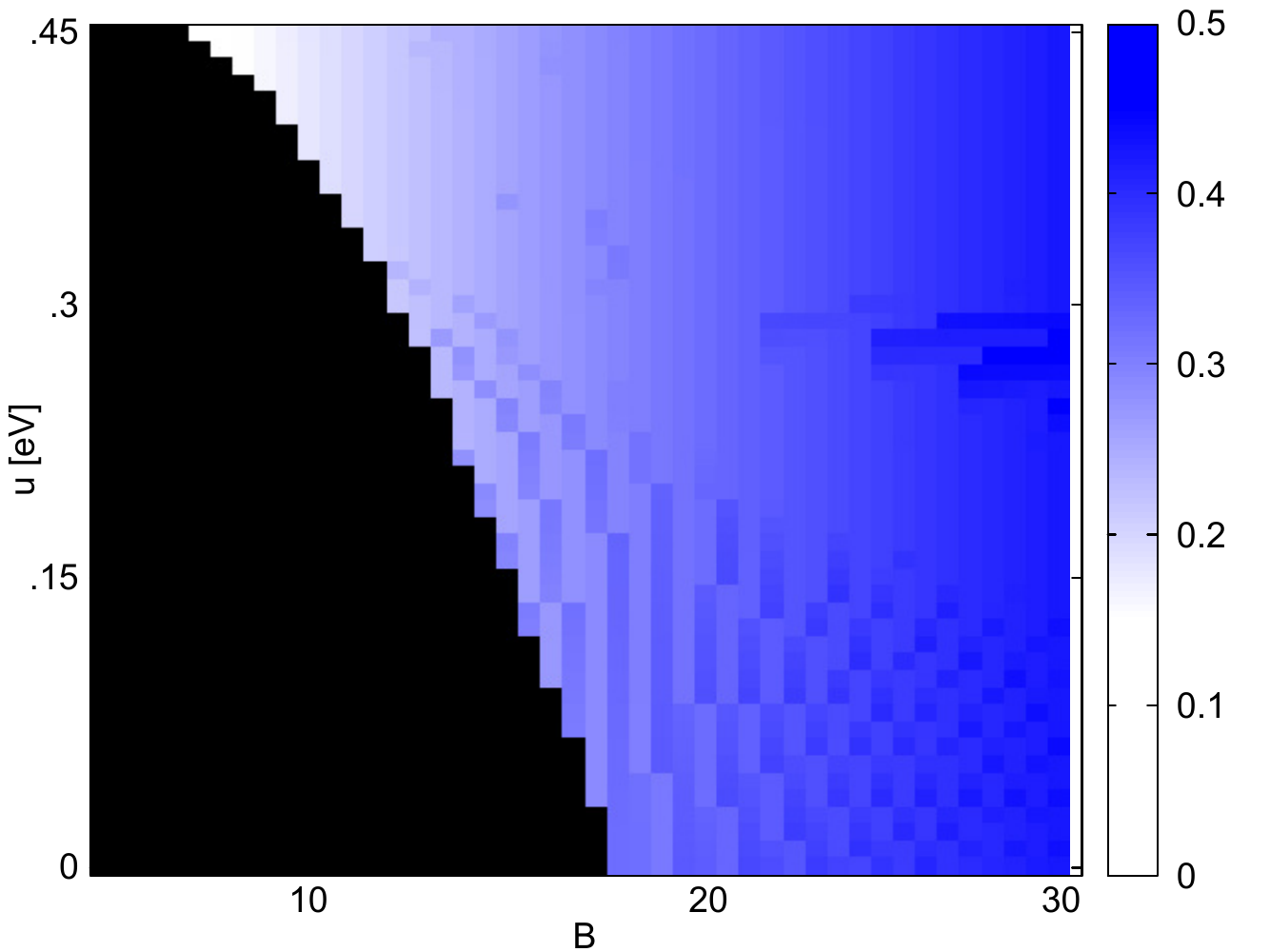}\\
	b. \includegraphics[width=.9\columnwidth]{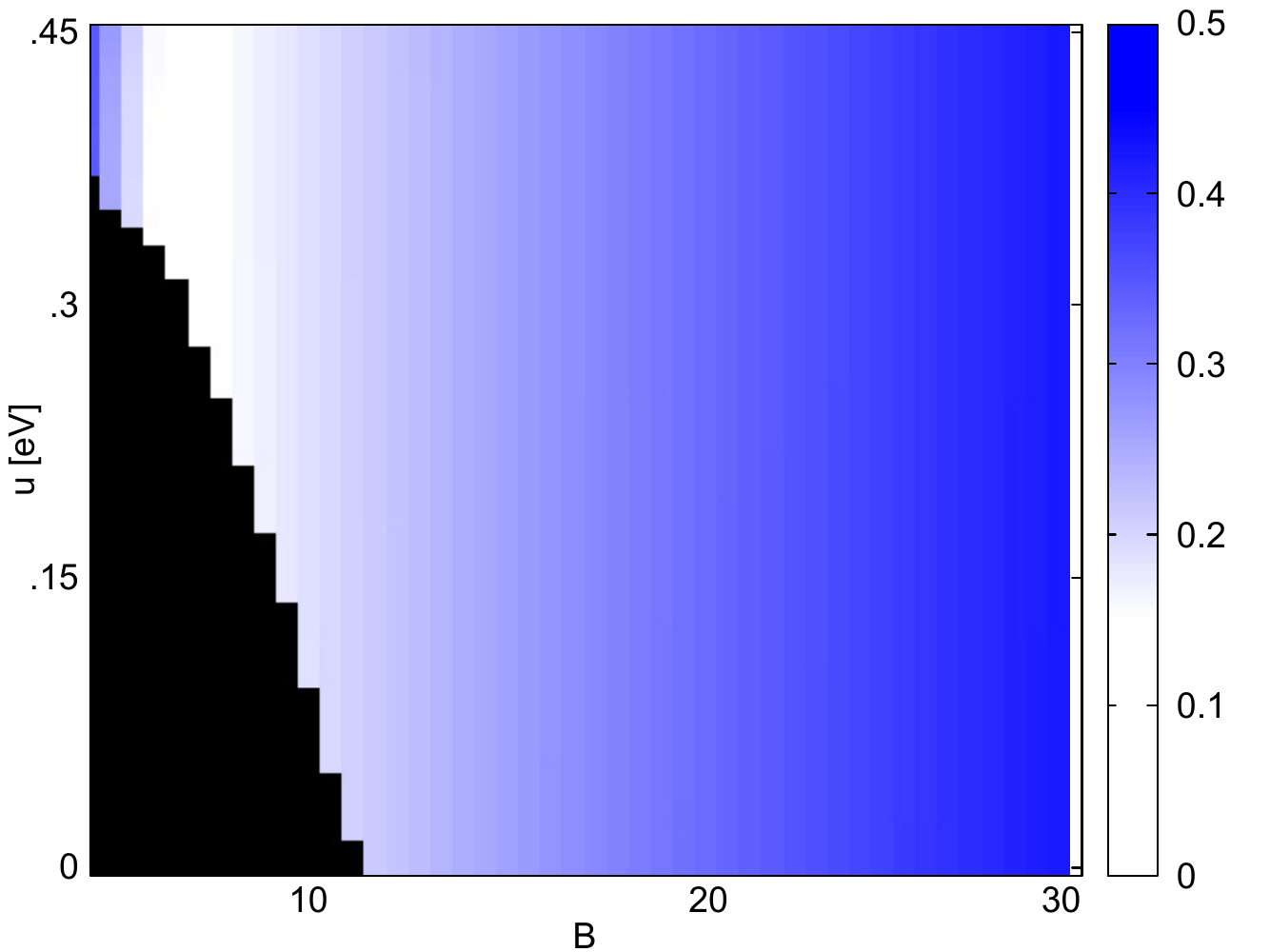}
	\caption{Weight outside the top layer first $N$ Landau levels for (a) $N=4$ and (b) $N=3$
	given by the color gradient, with the light region showing where the weight distribution
	may be favorable to non-Abelian states.
	The black region represents where the band gap above the degeneracy is less than the Coulomb
	energy.}
	\label{viability}
\end{figure}

\begin{figure*}[t]
\begin{center}
\includegraphics[width=.32\linewidth]{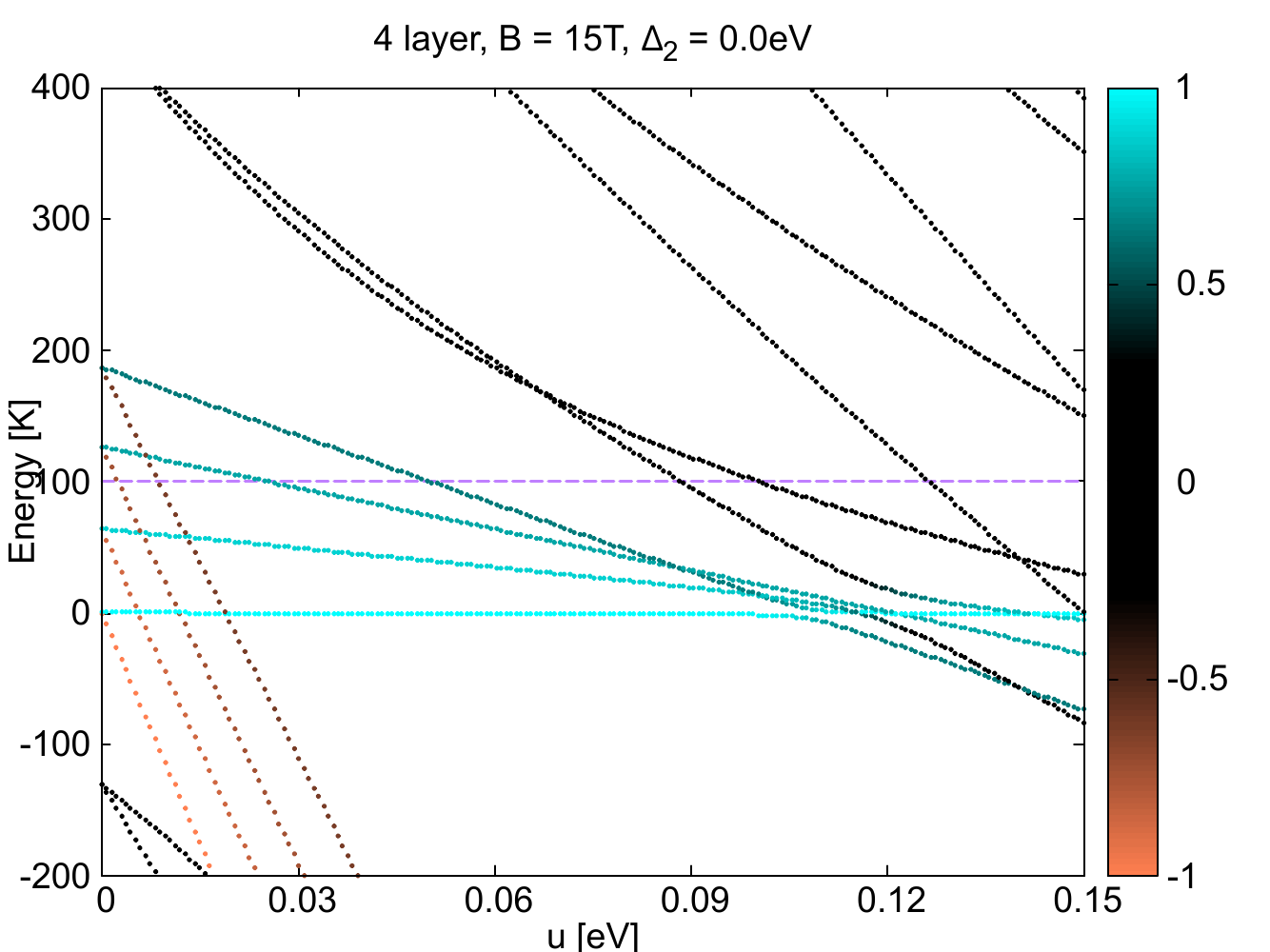}
\includegraphics[width=.32\linewidth]{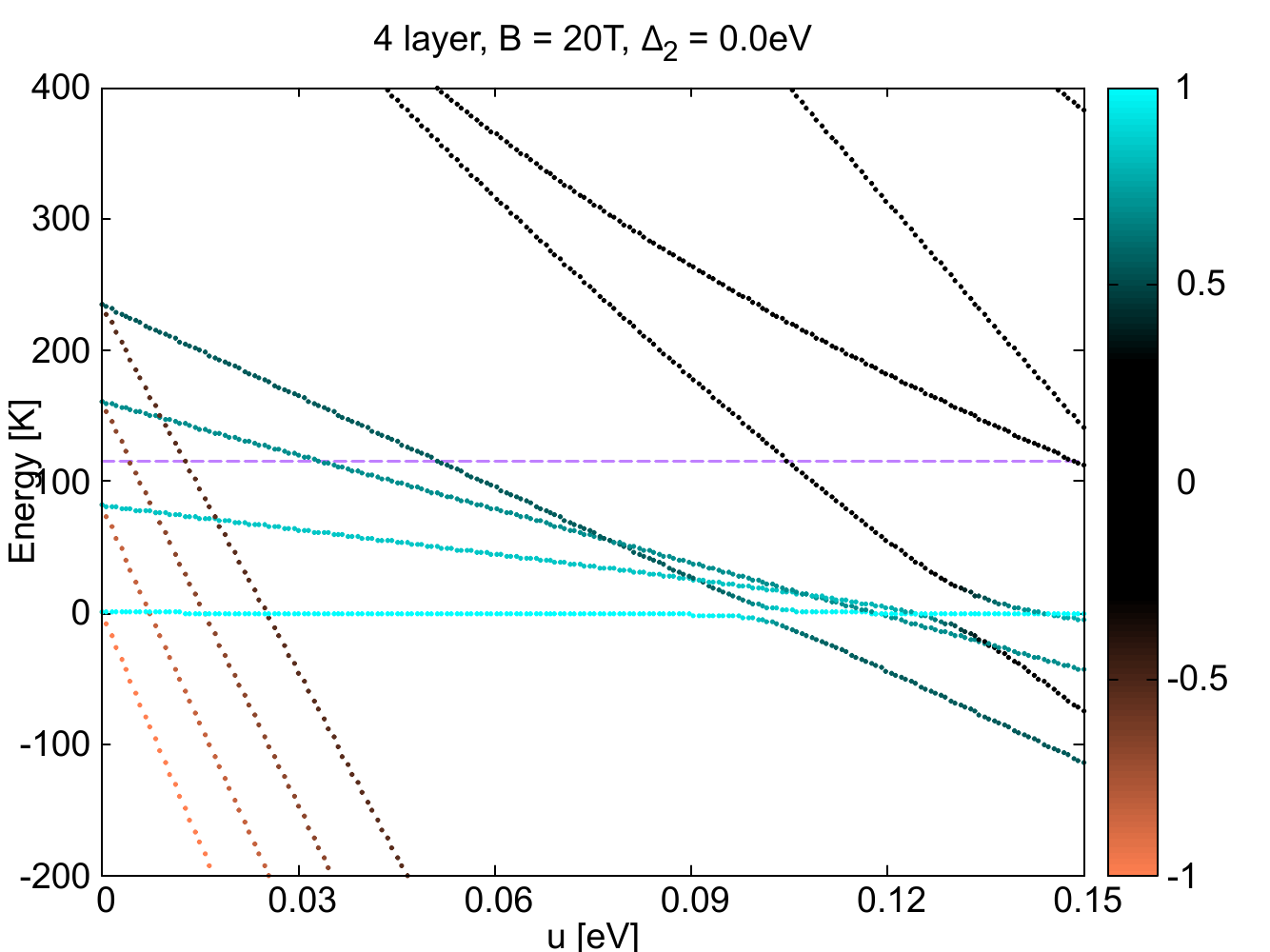}
\includegraphics[width=.32\linewidth]{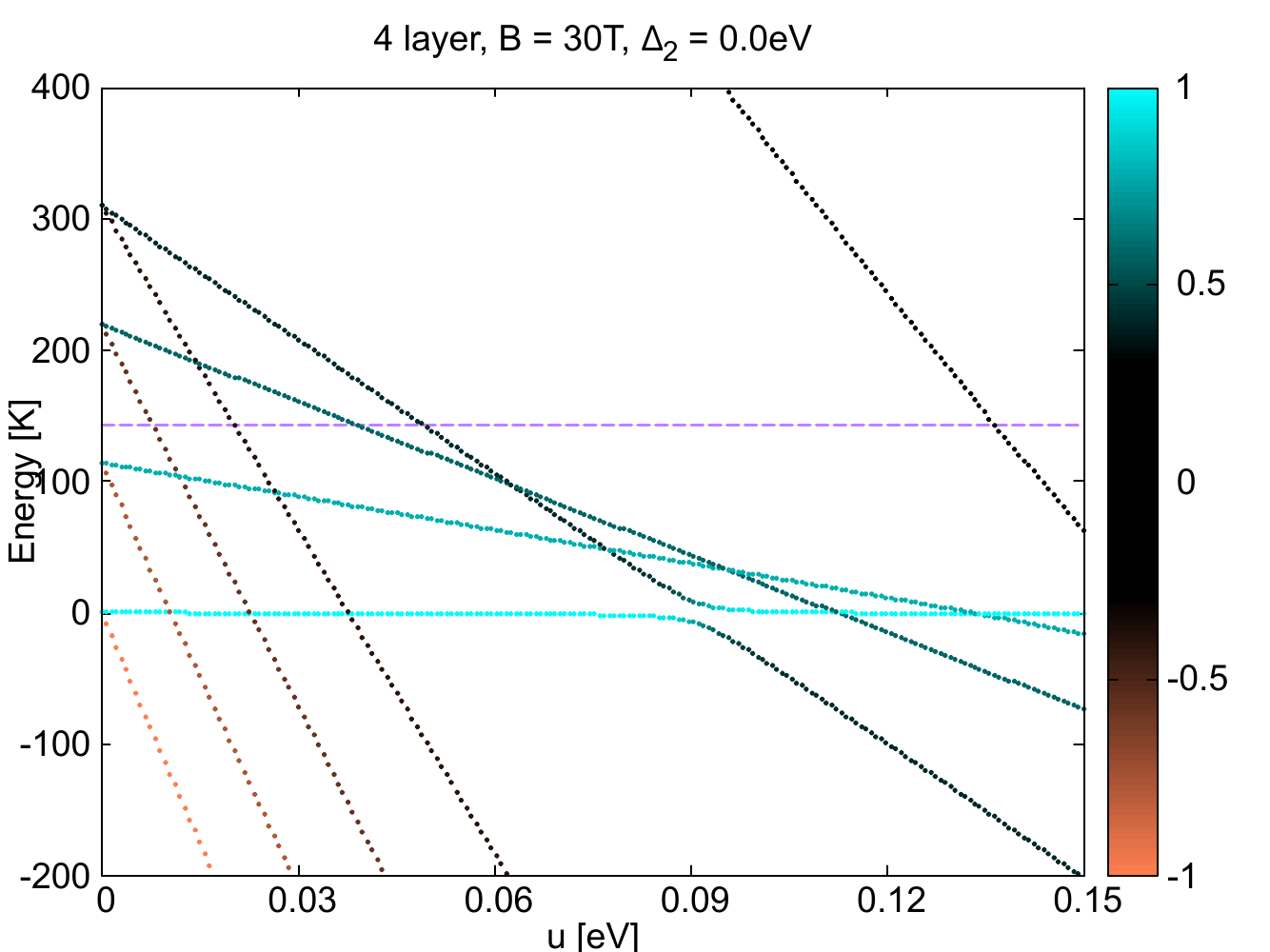}\\
\includegraphics[width=.32\linewidth]{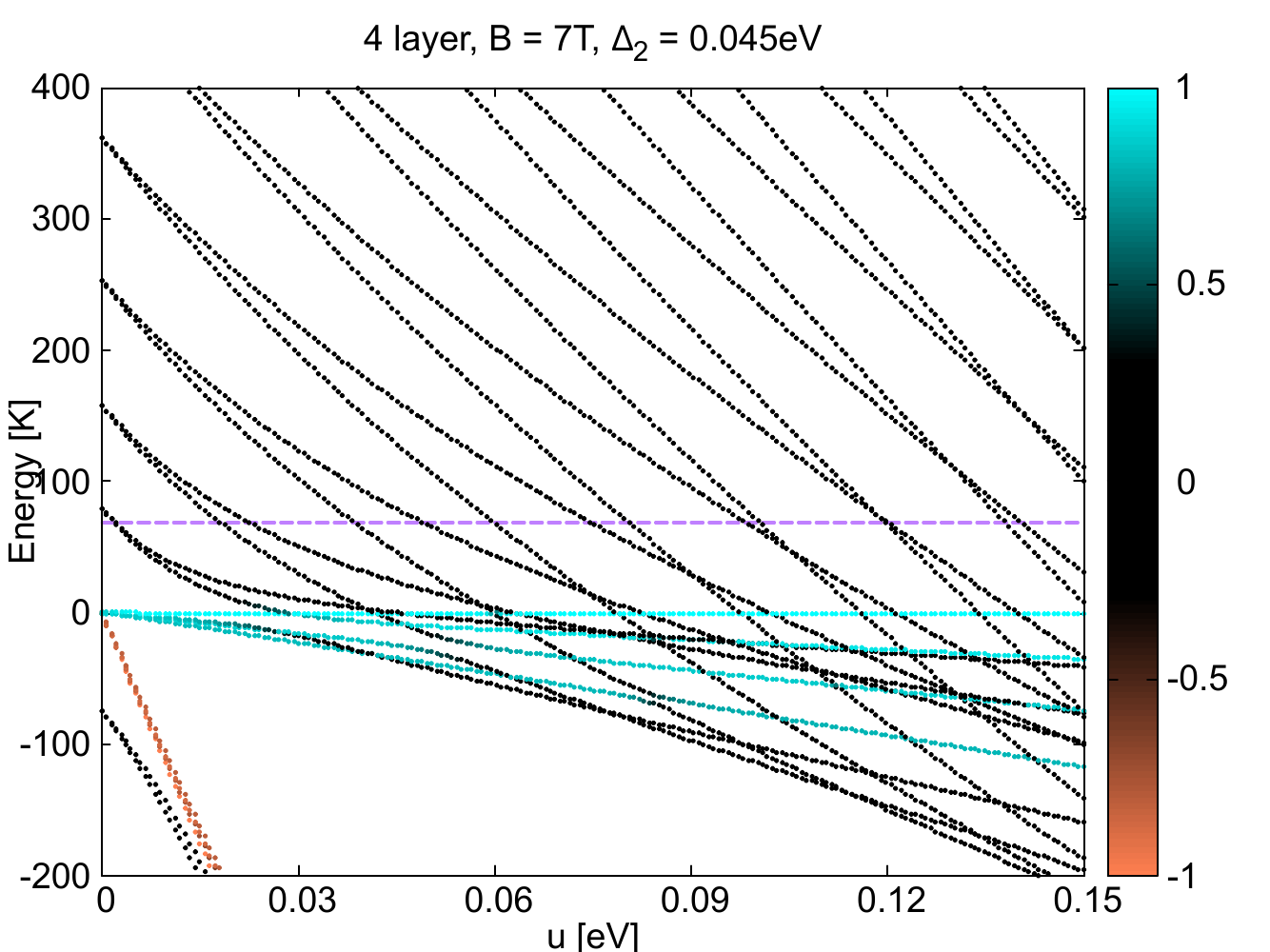}
\includegraphics[width=.32\linewidth]{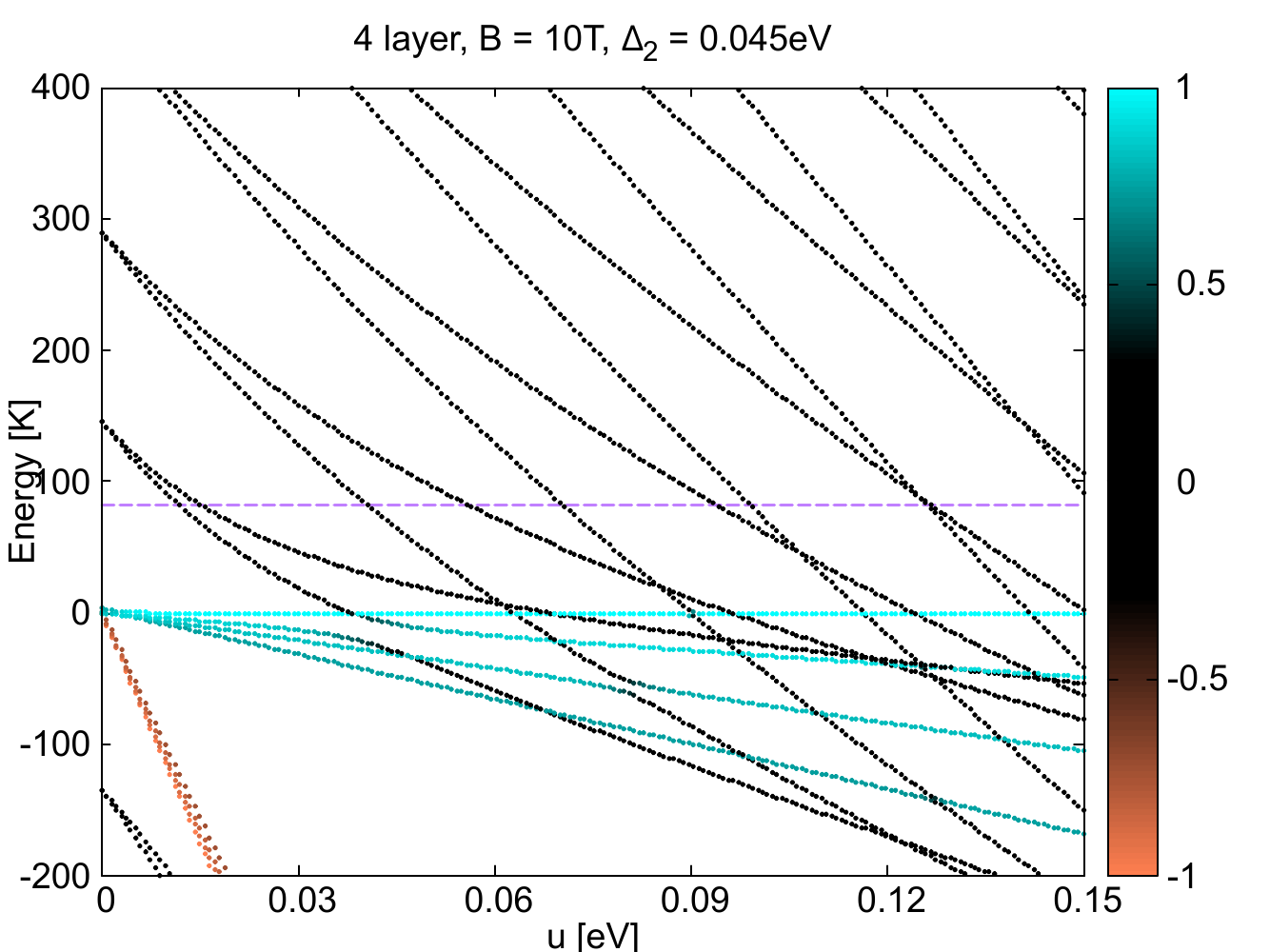}
\includegraphics[width=.32\linewidth]{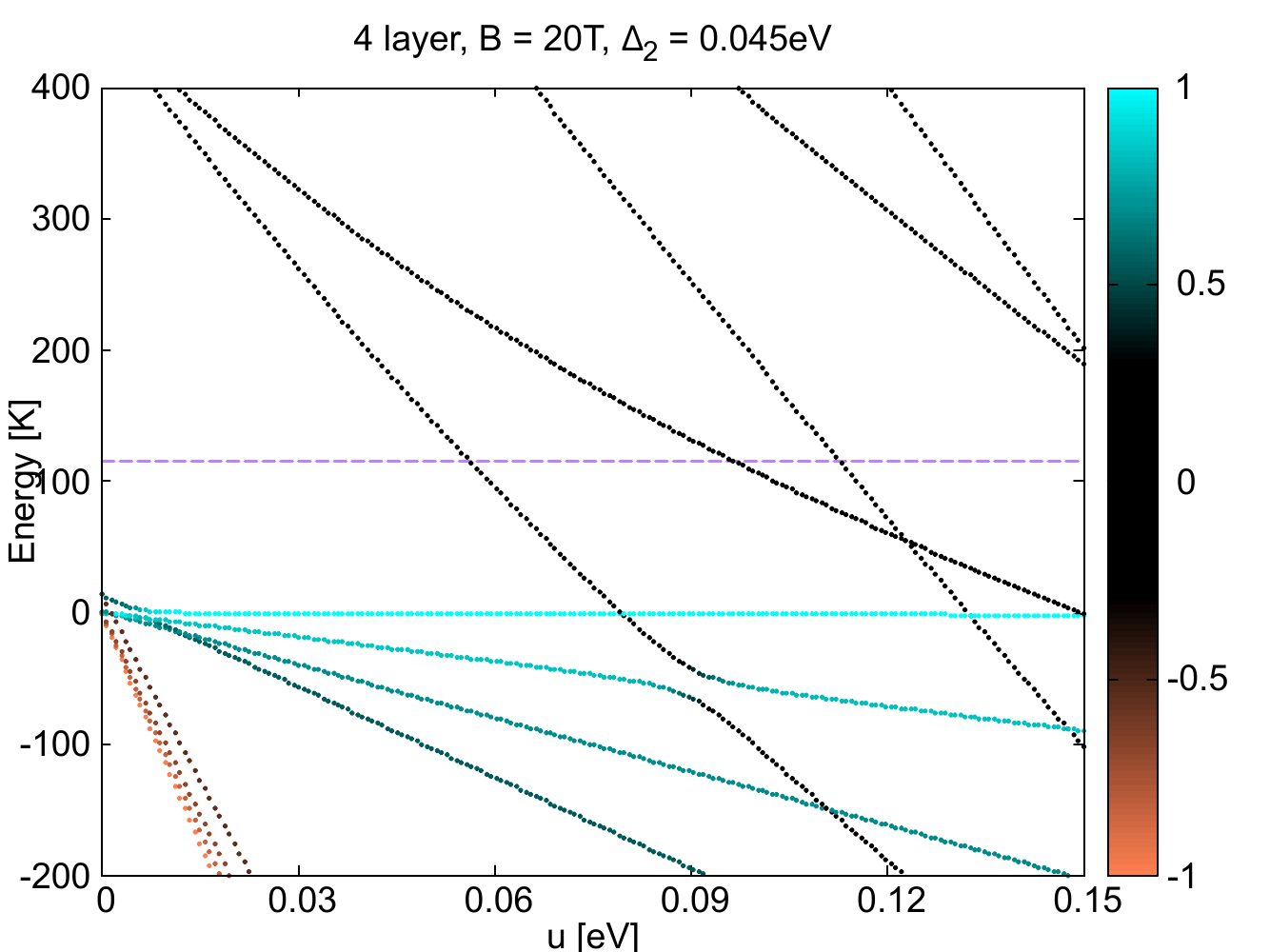}
\end{center}
\caption{ Landau levels in tetralayer rhombohedral graphene, from both $K$- and
$K'$-valleys).  The color indicates the weights of the wave function in the top
layer first 4 Landau levels for the $K$ valley (cyan, positive) or the bottom 
layer first 4 Landau levels for the $K'$ valley (orange, negative).  The 
spectrum is shown for $\Del_2 =0$ at $B=15,20,30$ Tesla and for
$\Del_2 =0.045$\,eV at $B=7,10,20$ Tesla.  The Landau level energy is in units
of Kelvin.  The horizontal axis is the displacement field $u$, the energy
difference between the top layer and bottom layer. The horizontal dashed line
marks the Coulomb energy $E_c$. }
\label{spec4} 
\end{figure*}

\begin{figure}[t]
\begin{center}
\includegraphics[scale=0.3]{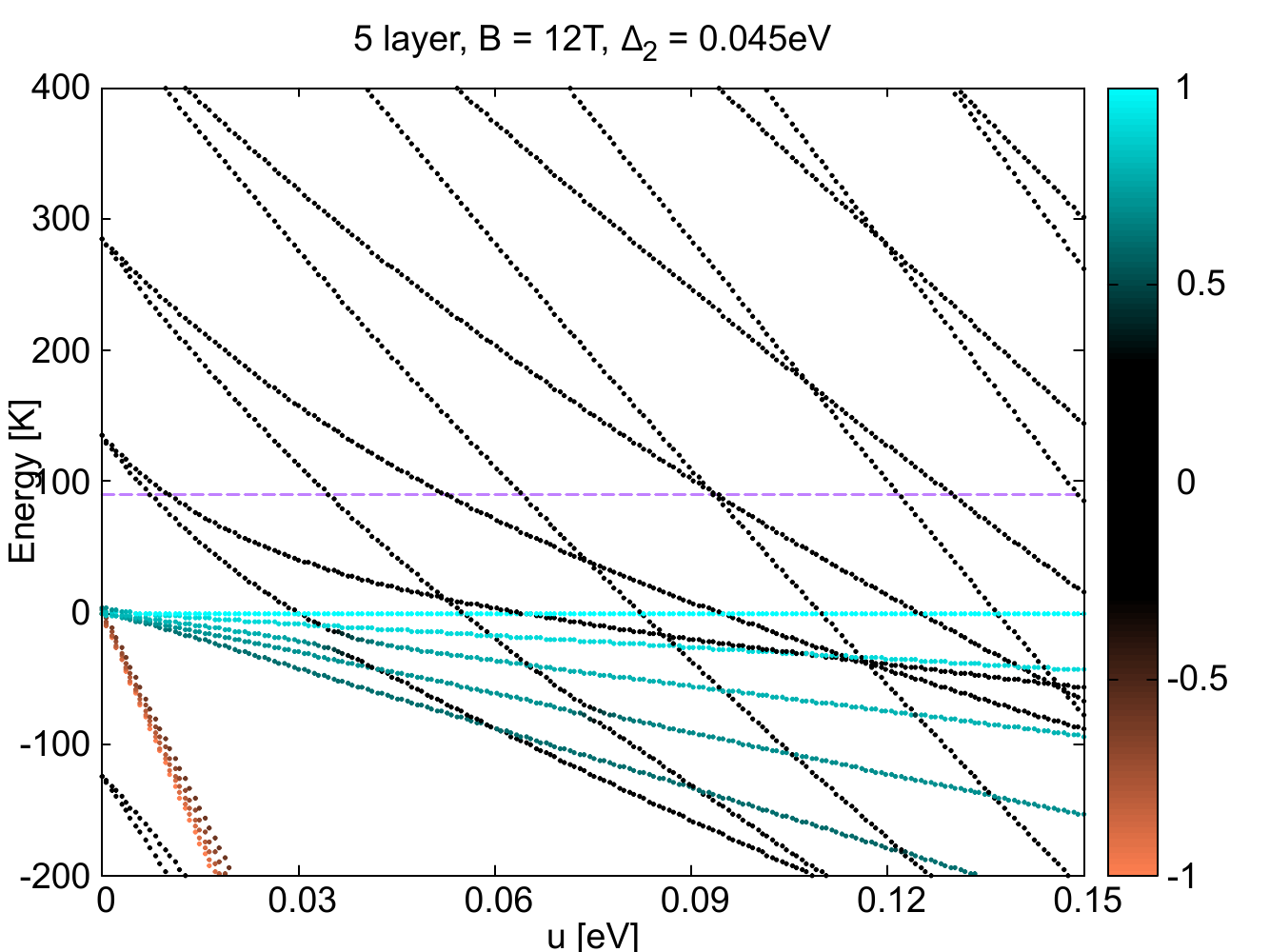}
\end{center}
\caption{ Landau levels from $K$- and $K'$-valleys 
for pentalayer rhombohedral graphene with $\Del_2 =0.045$\,eV at $B=12$ Tesla.
The color of the markers indicates the total weight of the Landau level in the
top layer first 5 Landau levels for the $K$ valley (cyan, positive) or the 
bottom layer first 5 Landau levels for the
$K'$ valley (orange, negative). The horizontal line marks the Coulomb energy.  }
\label{spec5} 
\end{figure}

\begin{figure*}[t]
\begin{center}
\includegraphics[width=.32\linewidth]{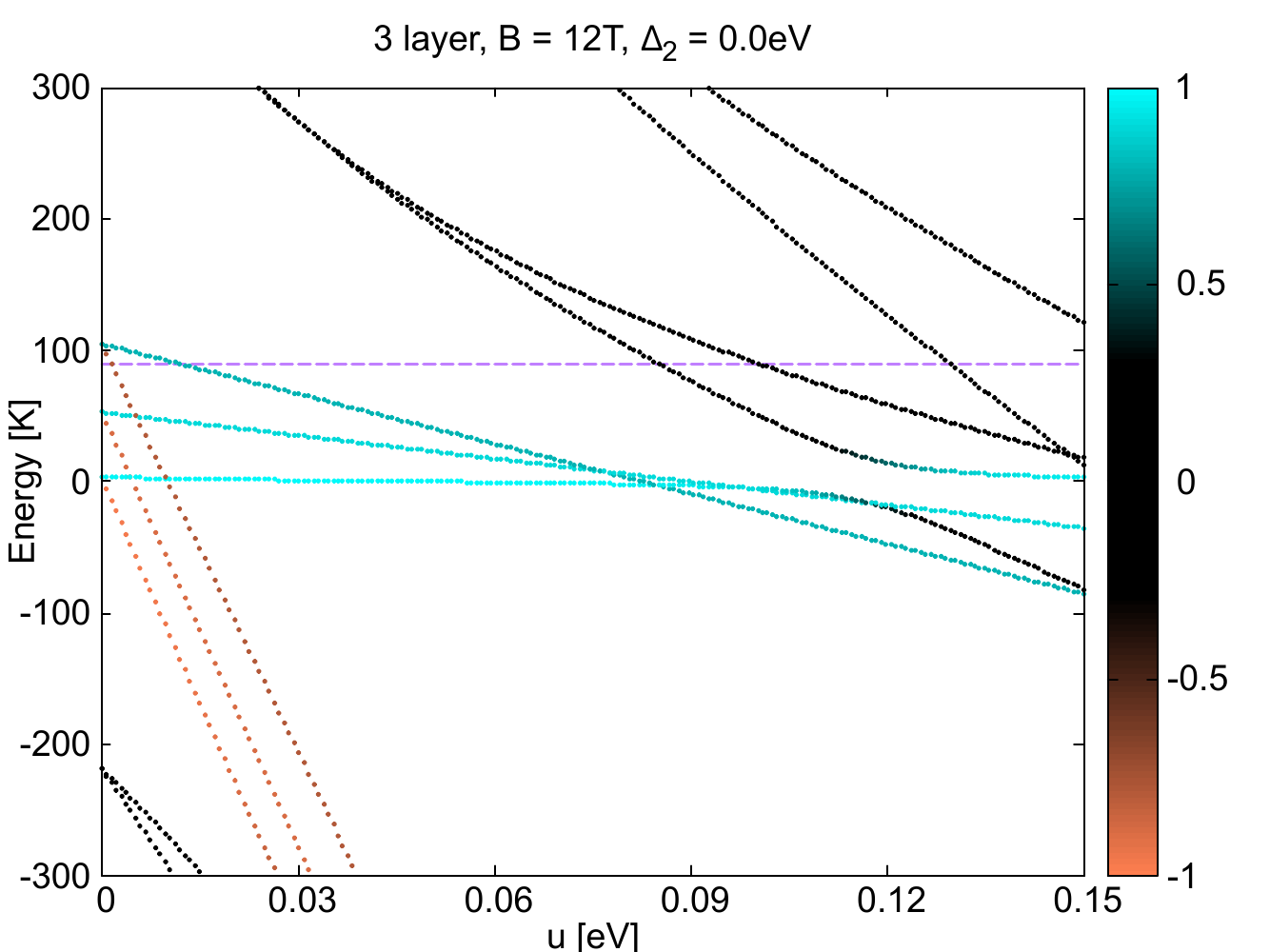}
\includegraphics[width=.32\linewidth]{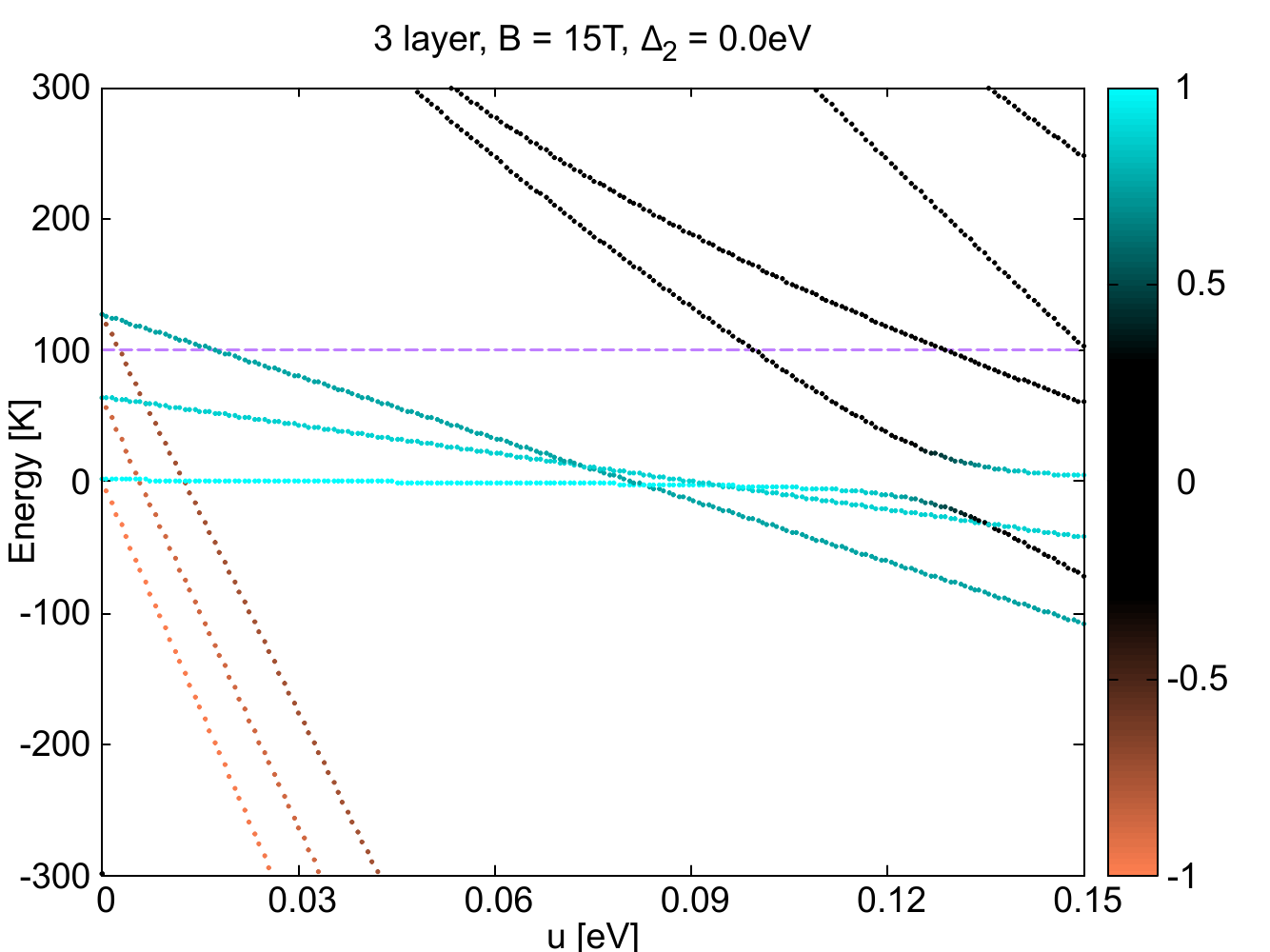}
\includegraphics[width=.32\linewidth]{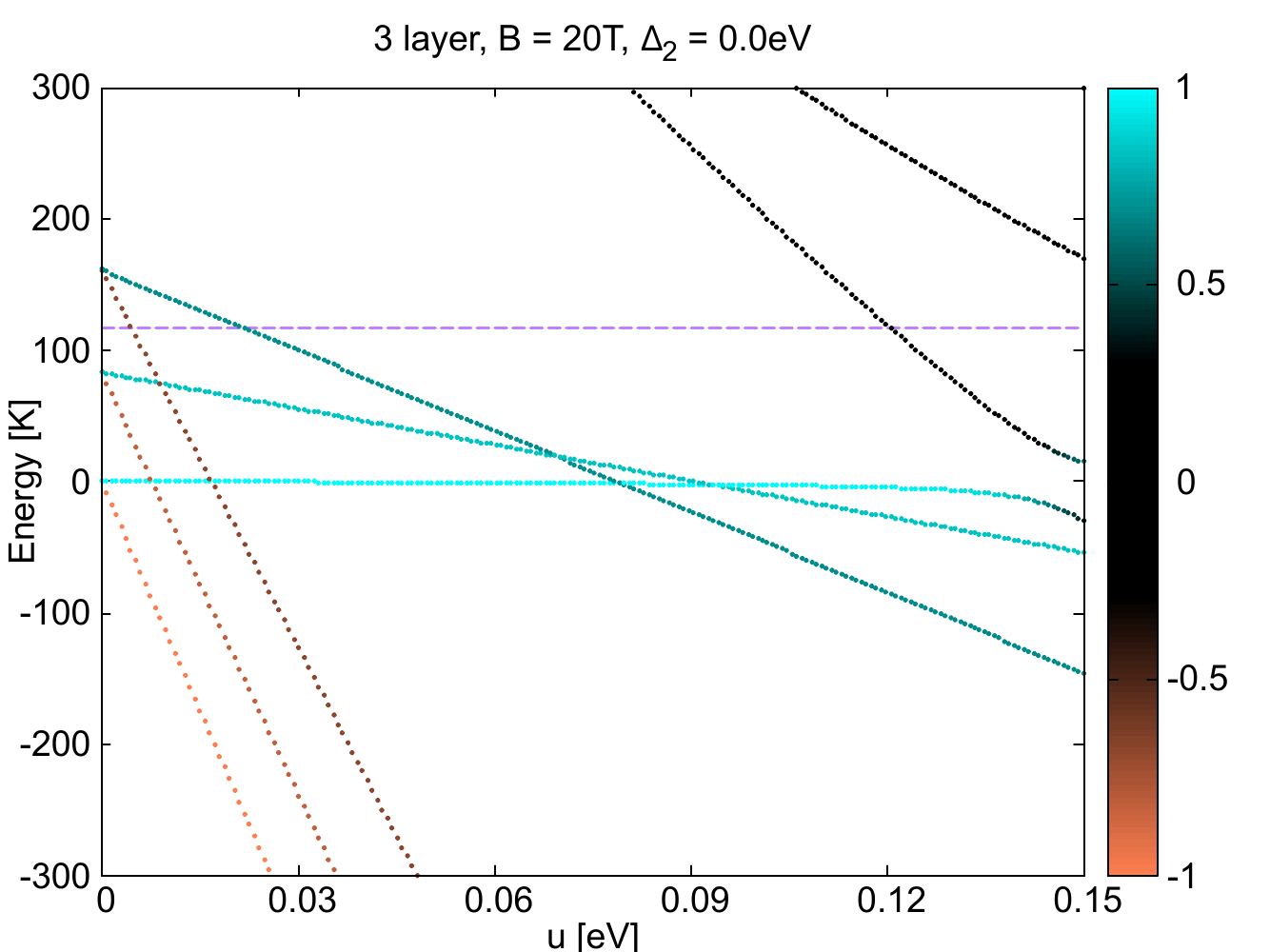}\\
\includegraphics[width=.32\linewidth]{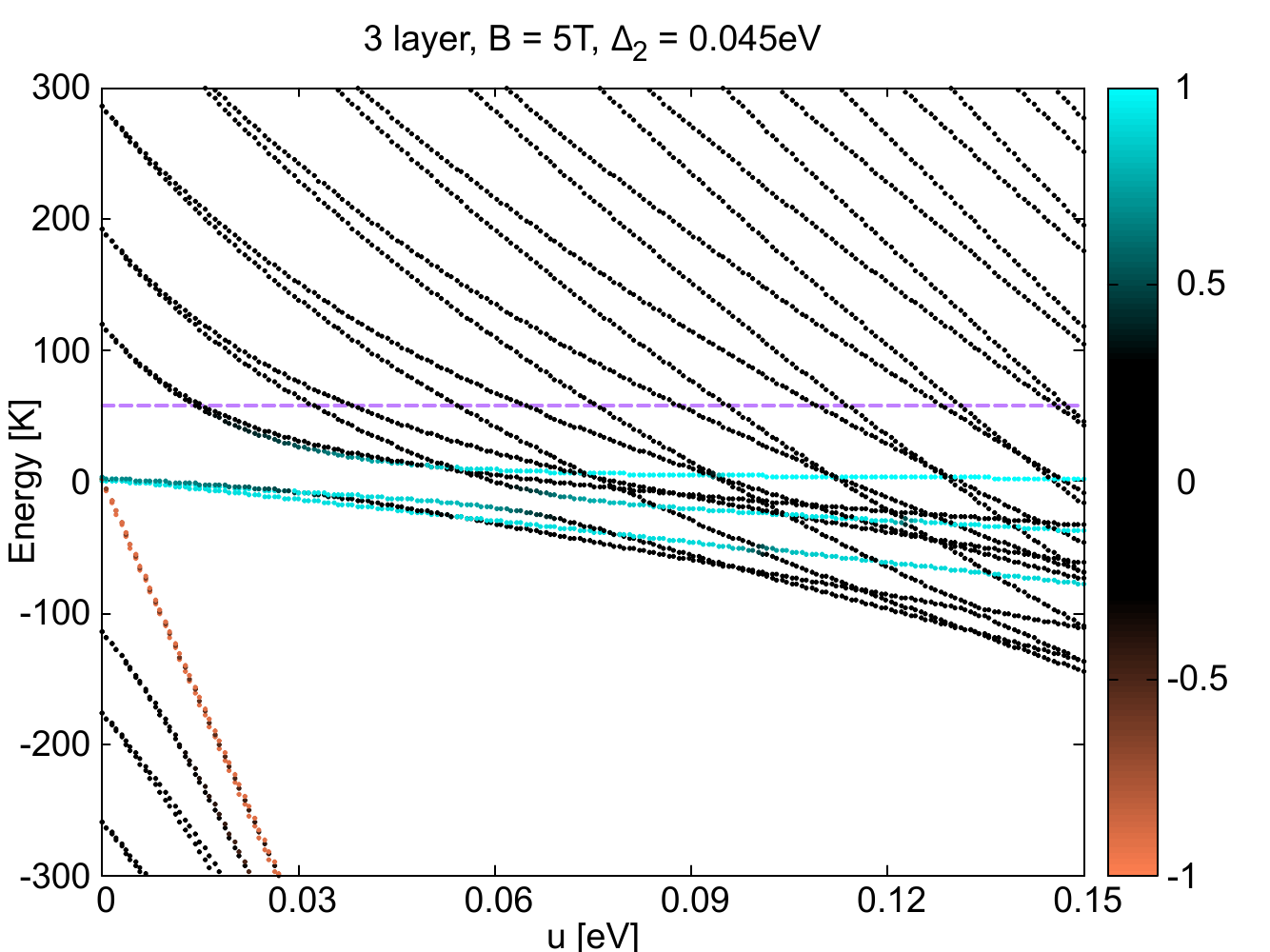}
\includegraphics[width=.32\linewidth]{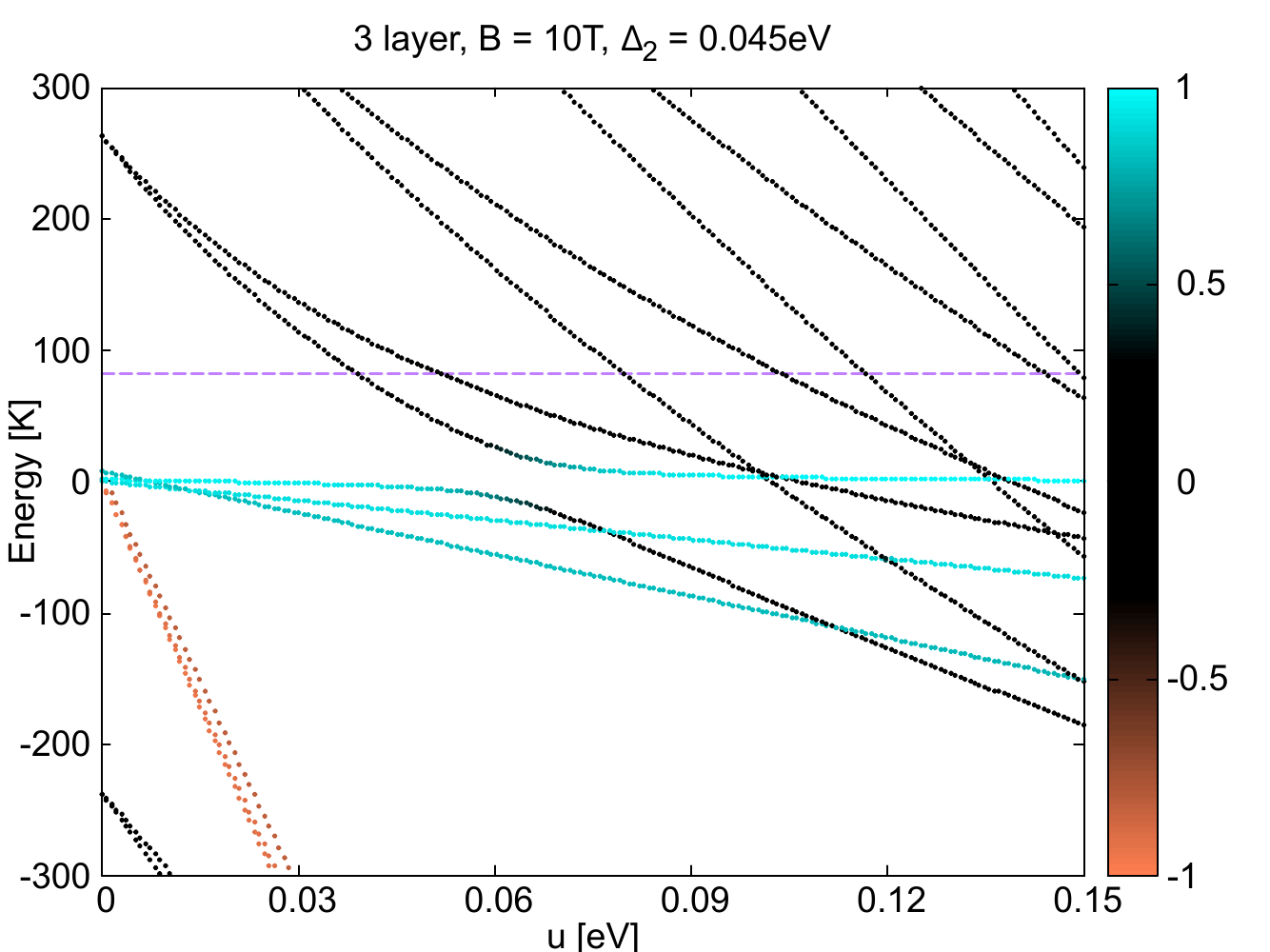}
\includegraphics[width=.32\linewidth]{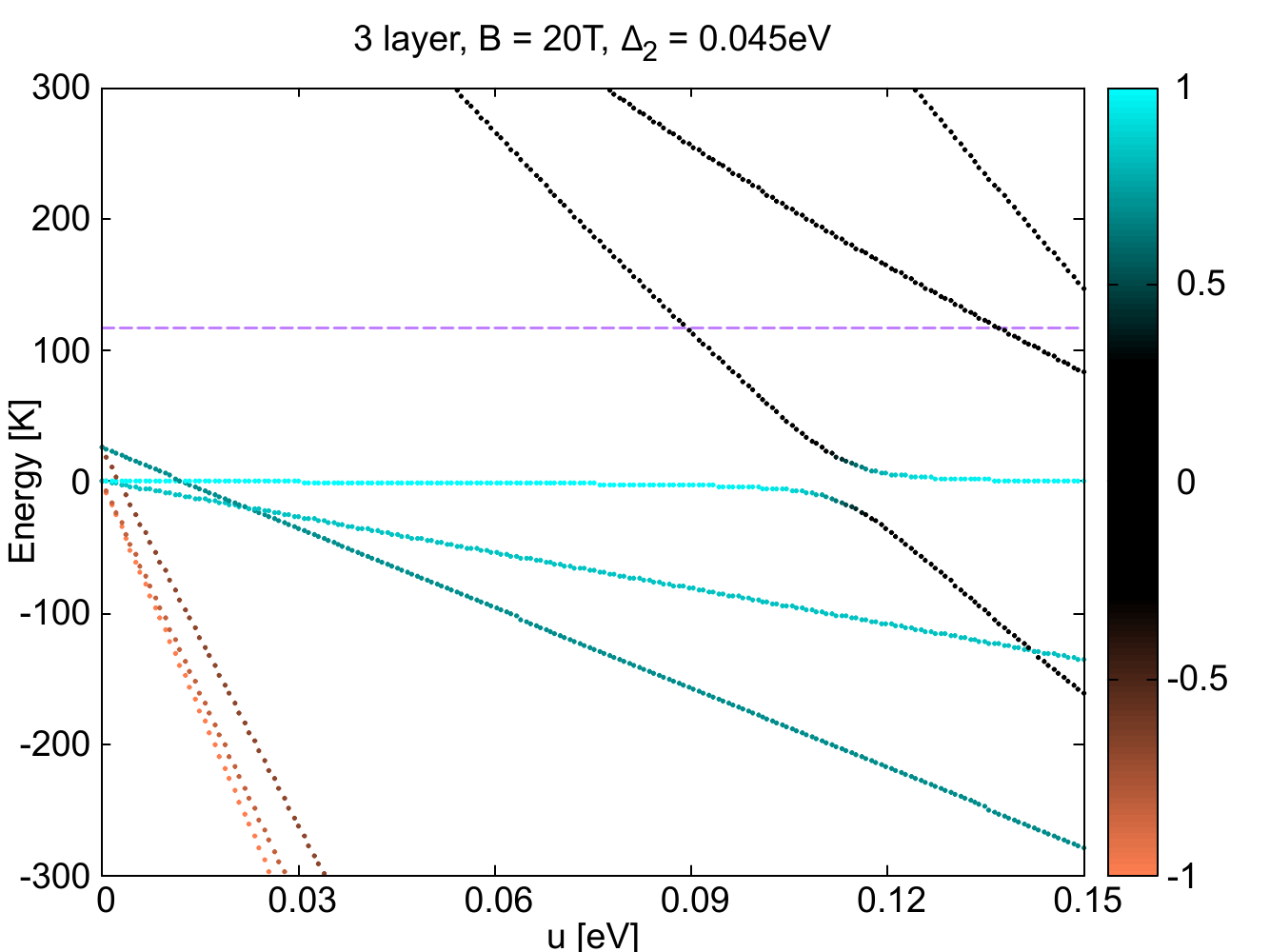}
\end{center}
\caption{ Landau levels in trilayer rhombohedral graphene, from the $K$- and
$K'$-valleys, for $\Del_2 =0$ at $B=12,15,20$ Tesla and for $\Del_2 =0.045$\,eV
at $B=1,10,20$ Tesla.  The Landau level energy is in units of Kelvin.  The
horizontal axis is the displacement field $u$, the energy difference between
top layer and bottom layer. The color indicates the total weight of the Landau
level in the top layer first three Landau levels for the $K$ valley (cyan, positive) 
or bottom layer first three Landau levels for
the $K'$ valley (orange, negative).  The horizontal line marks the Coulomb
energy $E_c$.  } \label{spec3} 
\end{figure*}

We now explore the effect of $\Delta_2$ on the Landau level degeneracy and estimate
appropriate ranges for the magnetic field. To do this, we compare both the degeneracy splitting
and the gap to higher bands against the Coulomb energy,
\begin{align}
 \frac{e^2}{\eps l_B} &= \frac{260}{\eps} 
\sqrt{B [\text{Tesla}] }\text{K}^\circ \equiv E_c,
\nonumber\\
l_B &=\sqrt{2\pi \hbar c/eB}=64.3/\sqrt{B[\mathrm{Tesla}]}\text{nm} .
\end{align}
where we take $\eps=10$.  We will consider band gaps above the Coulomb energy to be favorable.
When the gap is below the Coulomb energy, how it interferes in the filling of the degeneracy
will depend on details of the interaction, but for a conservative estimate we consider this
case to be non-ideal.  For the degeneracy splitting, we check that the splitting
is less than $30\%$ the Coulomb energy for favorability.

Examples of spectra for different $\Delta_2$ and $B$ are shown in Fig.~\ref{spec4}.  We find
that the degeneracy can be nearly cancelled by the displacement field except in cases where
the magentic field is large $\sim 30$ Tesla and $\Delta_2$ is small $\sim 0$ eV.  The 
band gap falls below the Coulomb energy for magnetic fields which are too small, $\sim 15$ Tesla
for $\Delta_2 = 0$.  The situation improves at $\Delta_2=0.045$ eV,
where the degeneracy is exactly cancelled and brought to zero displacement field, and the gap
can persist to lower magnetic field $\sim 7$ Tesla.

The intensity of color in Fig.~\ref{spec4} shows the weight in the top layer first four
Landau levels for the $K$ valley (cyan) and bottom layer first four Landau levels for the 
$K'$ valley (orange).  This is an estimate of the degree to which the single-species property of 
Eq.~\ref{singlespecies} is broken.  We will consider $<30\%$ of any band
outside this subspace to be favorable, $<40\%$ to be marginal, and $>40\%$ to be unfavorable.
A stronger estimate would be to take only the highest-weighted 
Landau level for each band; however, coherent mixture of bands may allow for independent 
tuning of each Landau level even when each band contains a mixture.  Thus we consider any distribution 
of weight in the first four Landau levels to be acceptable.  

Results of the favorability conditions are shown in Fig.~\ref{viability}(a), with the light region 
showing where all conditions are satisfied. The blue gradient represents orbital weight outside
the top layer first four Landau levels, and the black region shows where the band gap falls 
below the Coulomb energy. The degeneracy splitting is always less than $30\%$, usually less 
than $10\%$, wherever the other two conditions are satisfied, so we do not show this.
We find that the weight outside the first four Landau levels increases to the marginal case
($30-40\%$ for any one band) around 20 Tesla, and the gap drops below the Coulomb energy around
$B=18$ Tesla for $\Delta_2=0$ and $B=7$ Tesla for $\Delta_2 = 0.045$.

\subsection{Landau levels for trilayer rhombohedral graphene}

In Fig. \ref{spec3}, we present the Landau levels for trilayer rhombohedral
graphene with $\Del_2=0,\ 0.045$\,eV.  We see that the Landau levels from
$K$-valley are nearly degenerate for appropriate values of $u$, 
The energy gap to other Landau levels from the same valley is larger than the Coulomb energy
if $B>12$Tesla for $\Delta_2=0$, and for smaller magnetic fields if $\Delta_2=.045$ eV.  The
total weights on the interior layers are close to zero expect for the cases at
$B=20$ Tesla, where this exceeds 30\% for one band.  
When $\Delta_2=.045$ eV, we also find that the magnetic field must be $>5$ Tesla 
to keep weight within the first three Landau levels.  The favorability conditions are
summarized in Fig.~\ref{viability}(b), where again the light region shows conditions
favorable to non-Abelian states.  The orbital weight becomes marginal for realizing these
around 20 Tesla, and the gap falls below the Coulomb energy around 15 Tesla for $\Delta_2=0$.
For $\Delta_2=0.045$ eV, orbital weight outside the first three Landau levels sharply increases
below 5 Tesla.  This allows us to conclude that non-Abelian
FQH state can be realized in trilayer rhombohedral graphene, for a magnetic
field of 12 -- 20 Tesla when $\Delta_2 = 0$, or 5 -- 20 Tesla when $\Delta_2 =
45$ meV.

\subsection{Landau levels for pentalayer rhombohedral graphene}

For pentalayer graphene, we find only a marginal case for realizing non-Abelian states, 
 with $\Del_2=0.045$\,eV and $B=12$ Tesla shown in Fig.~\ref{spec5}.  We see that the
Landau levels from $K$-valley are almost exactly degenerate at $u=0$, whose
energy gap to other Landau levels from the same valley are larger than the
Coulomb energy.  The total weights on the interior layers are $0.15,0.05,0.20,0.25,0.35$
for the five degenerate Landau levels, which is marginal by our criteria.  

\section{Interacting model}

\begin{figure}[t]
\begin{center}
\includegraphics[width=\columnwidth]{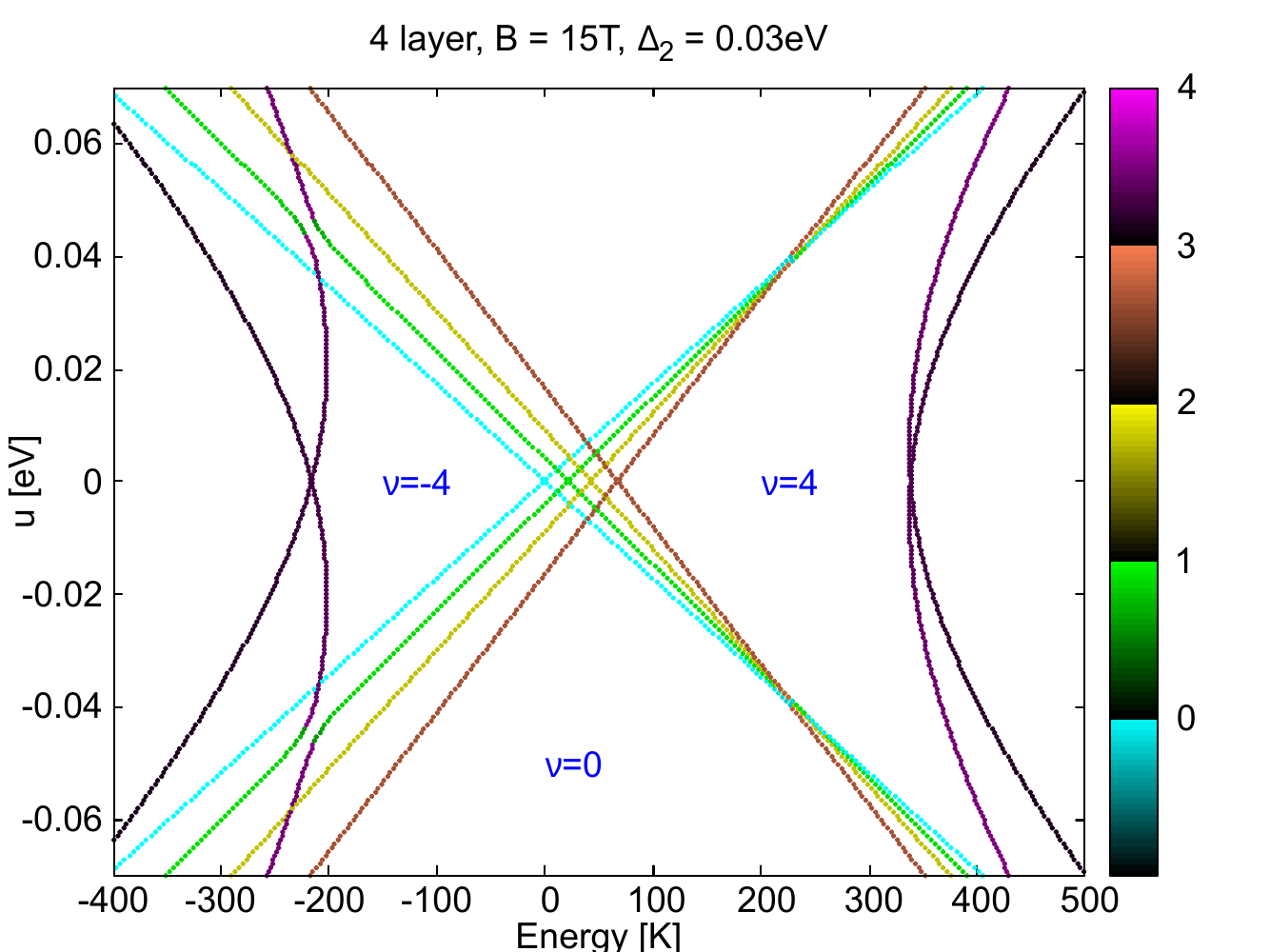}
\end{center}
\caption{Band structure including both $K$ and $K'$ valleys (but not electron 
spins) for four layers of graphene at $B=15$ Tesla and $\Delta_2=.03$ eV.  The 
$x$-axis represents energy and the $y$-axis displacement field, to compare with 
the interacting diagrams.  The color indicates the highest-weighted Landau 
level in the band, with shading based on how much weight is in the top (or 
bottom for $K'$) layer for that band, from dark (little weight) to bright 
(concentrated weight).  The Coulomb energy is about 100 Kelvin.
We set zero filling as the mid-point between Hall conductance $-4$ (all low bands empty)
and Hall conductance $4$ (all low bands full), marked in blue text.}
\label{phase4} 
\end{figure}

\begin{figure*}[t]
\hfil	(a) \includegraphics[width=3.0in]{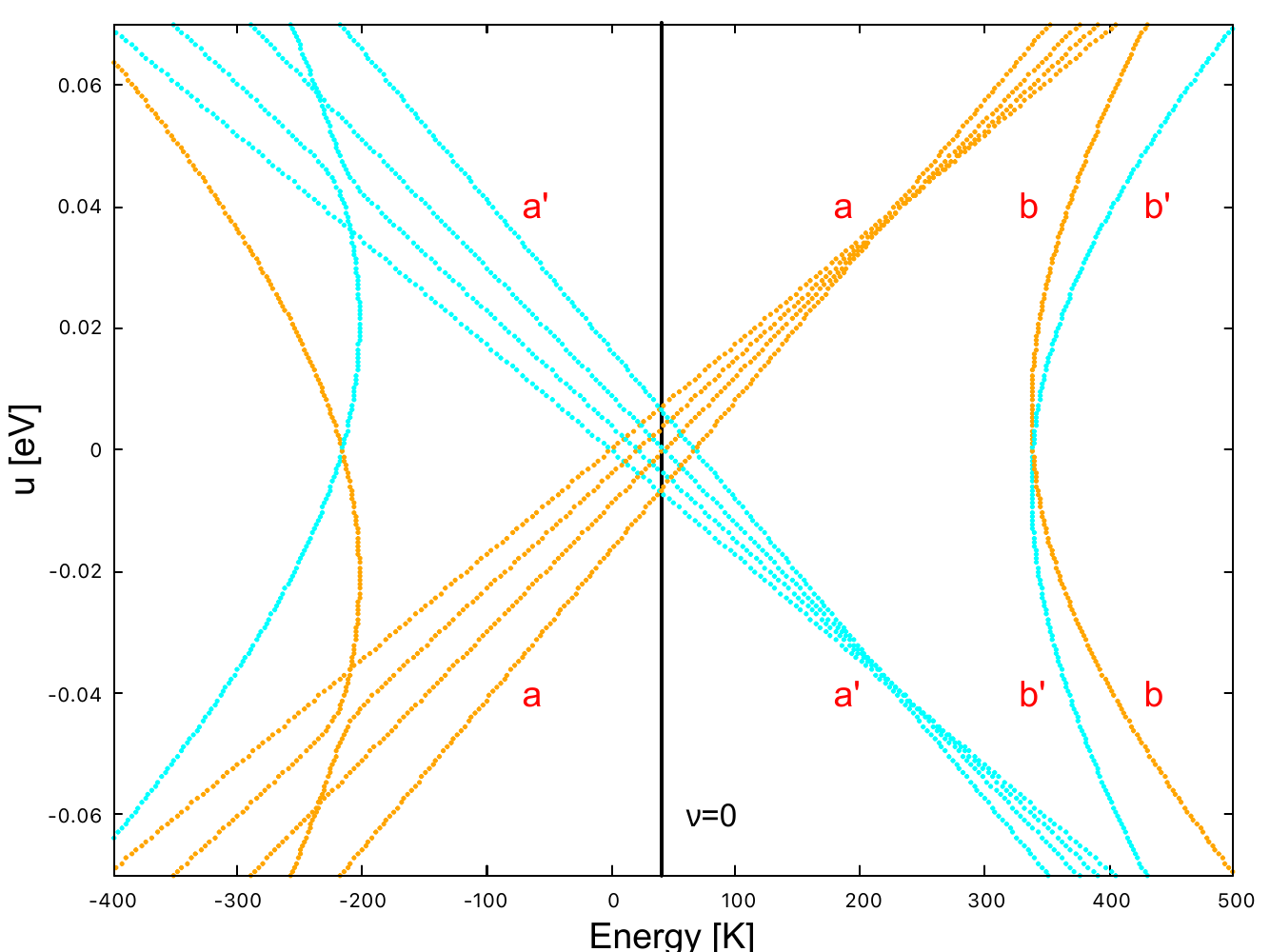}
\hfil	(b) \includegraphics[width=3.0in]{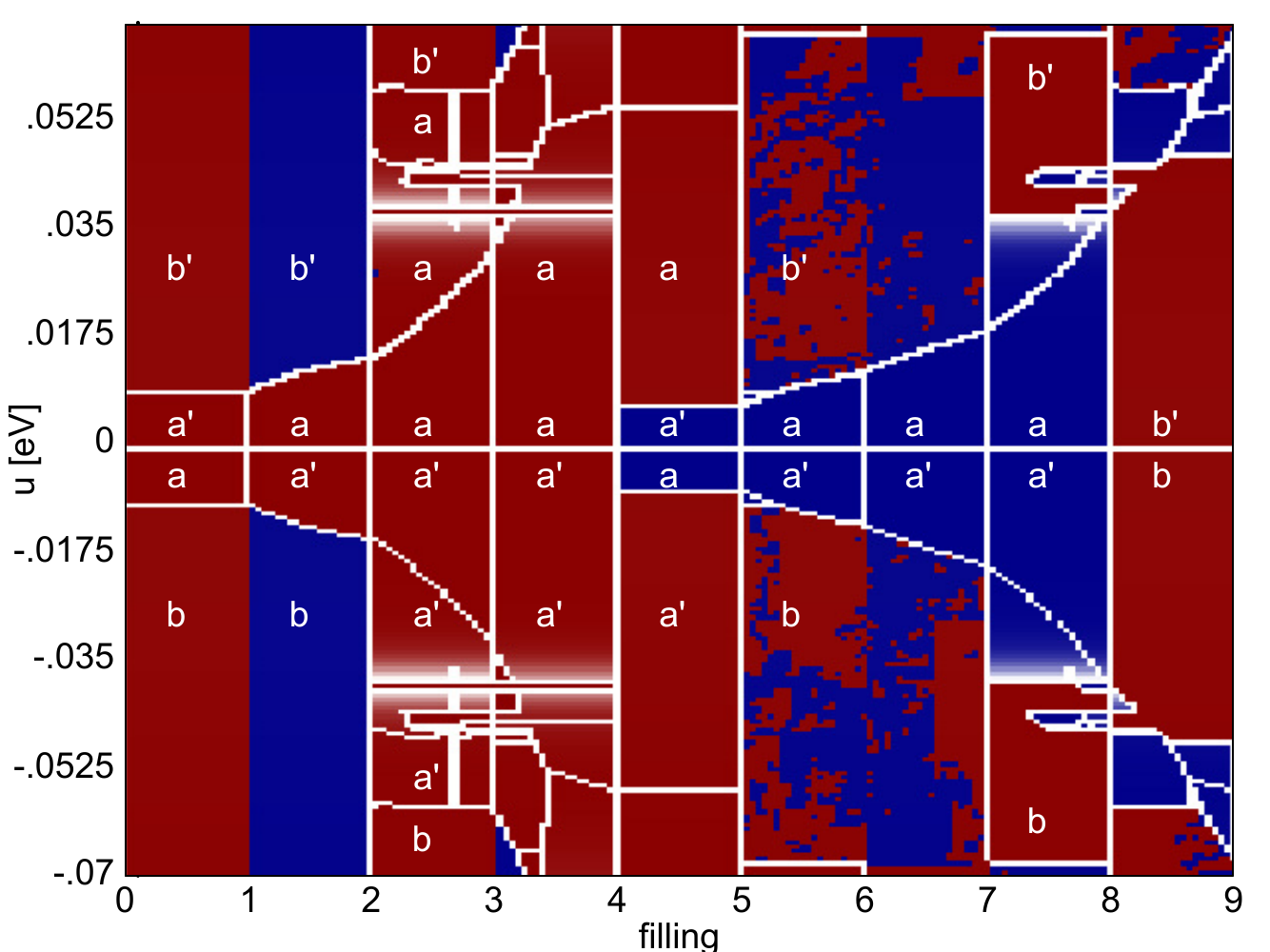}
\\
\hfil	(c) \includegraphics[width=3.0in]{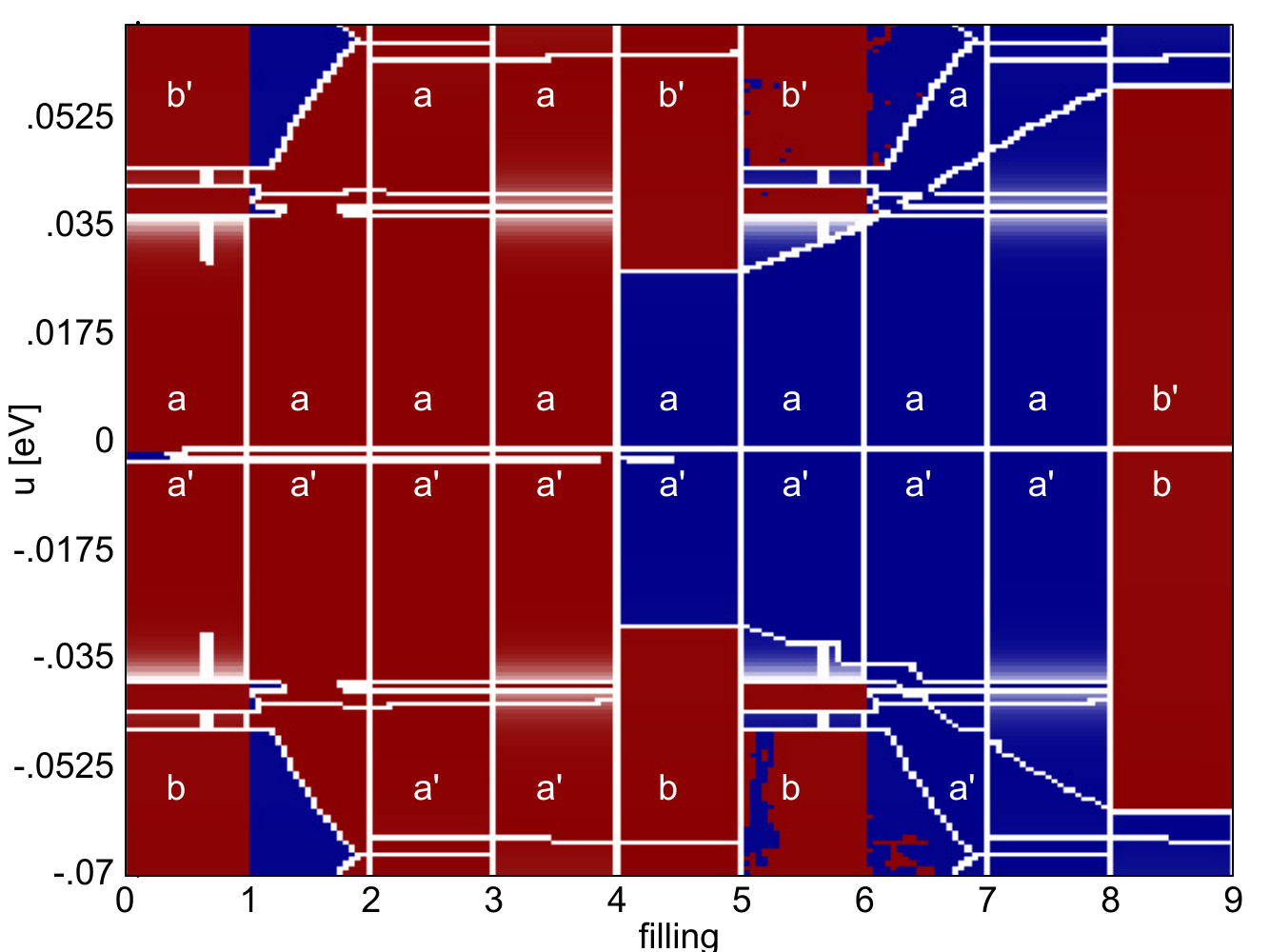}
\hfil	(d) \includegraphics[width=3.0in]{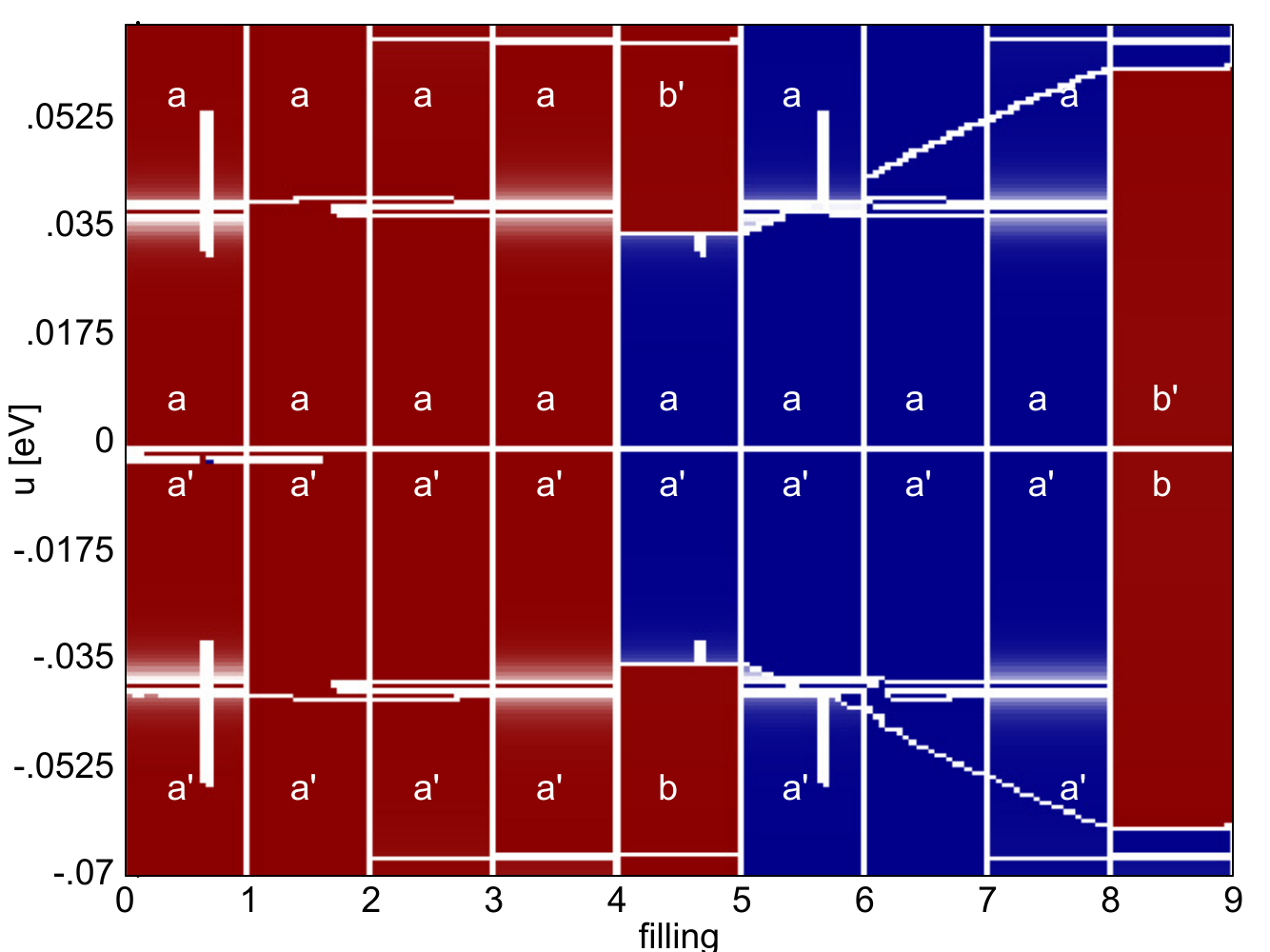}
	\caption{(a): Reference band structure for tetra-layer graphene at 15T
magnetic field, with valleys distinguished by color.  The $a$, $b$, $a'$, $b'$
labels correspond to: $a$ for the four $K$-valley bands
near zero energy concentrated near the top surface of the tetra-layer graphene.
$a'$ for the four $K'$-valley bands near zero energy concentrated near the
bottom surface.  $b$ for higher energy $K$ bands near the top surface.  $b'$
for higher energy $K'$ bands near the bottom surface.  (b-d): Wavefunction
change as a function of filling for interaction strength (b) 100 Kelvin, (c) 60
Kelvin, (d) 50 Kelvin.  White lines mark changes in Landau level index of the
active bands, and red/blue ($a,b,a',b'$) indicate spin-up/down ($a,b,a',b'$
bands) is being populated.  The filling $\nu$ is measured from charge
neutrality point, also labeled in plot (a). The vertical white bars are
the $SU(3)_2$ Fibonacci states.
} \label{wf_change}
\end{figure*}

In order to approximate the spin dynamics and experimental signatures of
the Landau level degeneracy, we now build a simple model to simulate an
exchange effect.  We will minimize the energy
\begin{align}
	\tilde E = \sum_{i,j,\sigma,\sigma'} E_i\nu^\sigma_i + (\nu_i^\sigma - 1/2) V^{\sigma\sigma'}_{ij} (\nu_j^{\sigma'} - 1/2)
\end{align}
where $\nu_i$ is the filling of the $i$th band, $0\leq \nu_i \leq 1$ and
$\sigma$ labels the spin-valley isospin.  The interaction energy we define as
\begin{align}
	V_{ij}^{\sigma\sigma'} = -|V| \frac{v_i \cdot v_j}{|\ell_i - \ell_j|+1} \delta_{\sigma\sigma'}
\end{align}
where $v_i$ is an $N$-component spinor specifying the orbital weight on each
graphene layer, and $\ell_i$ is the maximally weighted Landau level index for
the $i$th band.  $|V|$ is the overall energy scale, i.e. the Coulomb energy.  

This interaction quantifies the energy gain by an antisymmetrized state over an
uncorrelated state.  For states of different spin/valley quantum number, there
is no energy gain, so $V_{ij}^{\sigma\sigma'}$ is zero when $\si\neq \si'$.
For states of the same spin/valley quantum number, there will be a decrease in
energy from antisymmetrizing, driving the exchange effect.  This will depend on
how much orbital weight is in the same layer for the two states, captured by
the layer index spinor dot product $v_i \cdot v_j$.  It will also depend in
some complicated way on the difference in Landau level index.  Since the
orbital overlap of an uncorrelated state should decrease with Landau level
index based on differing orbital profiles, we model this using
$1/(|\ell_i-\ell_j|+1)$ as a simple approximation. 

We now focus on the case of 4-layer graphene at 15T and $\Delta_2=.03eV$.
The bands are plotted in Fig.~\ref{phase4}, with the dominant Landau level
and corresponding weight indicated by color, to give an idea of the structure of $V_{ij}$.
The valley that each band originates in is marked in Fig~\ref{wf_change}(a).
Performing the minimization of $\nu_i$ for each total filling and displacement
field at different interaction strengths $V$, we find the diagrams in
Fig.~\ref{wf_change} (b-d), with $V=100,60,50$ kelvin respectively. 

Zero filling is defined as follows.  In general, the smallest fillings in 
Fig.~\ref{wf_change} begin with the lower bands (labeled $a$ for negative 
displacement field or $a'$ for positive displacement fields) totally filled by 
both spins.  Close enough to zero displacement field, if we were to decrease 
the filling by 4, we would reach a case with these 4 lower bands empty for one 
spin.  Likewise if we increase the filling by 4, we expect to fill 4 low bands 
from the other valley for one spin ($a'$ for negative displacement field or $a$ 
for positive displacement field).  In both cases, the filling of the other spin 
remains fixed.  The two situations, with the near-zero bands totally filled or 
totally empty for one spin, correspond to a change of 8\ in the Hall 
conductance, and they are related by particle-hole symmetry.  Thus we argue the 
filling halfway between these cases should be zero Hall conductance, which we 
label $\nu=0$.  For strong interactions, the filling pattern may become more
complicated as we will see, but in general we find a filling with the lower 
bands totally filled by up and down spins which we call $\nu=0$.

In Fig.~\ref{wf_change}, white lines represent a change in the Landau level 
index of the actively filled band, quantified by the change of the maximally
weighted Landau level plus the weight in that Landau level, so we observe
abrupt changes when the index shifts as well as some gradual changes as the
weight changes smoothly.  The vertical white lines at integer fillings indicate
that we start to fill a new Landau level after filling the previous one. 
Abrupt changes in the Landau level index indicated by horizontal white lines
correspond to first-order phase transitions (\ie Landau level crossings for
the non-interacting case) induced by changing displacement field, which  should be
apparent in experiment.  

Red and blue indicate the change in total spin magnetic moment $M$ 
as a function of filling $\nu$, $\frac{\dd M}{\dd \nu} > 0$ for red and 
$\frac{\dd M}{\dd \nu} < 0$ for blue, which is also experimentally
measurable.  Horizontal white lines around displacement field $\pm .04$ are a
good indication of the band degeneracy.

To understand consequences of the valley-exchange effect, it is useful to 
consider the valley that each band originates from, indicated in
Fig.~\ref{wf_change}(a).  In all diagrams Figure~\ref{wf_change}(b-d), we find 
a tendency to valley polarize, especially near zero displacement field where 
there is a single transition as
the active bands switch from one valley-polarized set to the other.  Depending
on the interaction strength, there can be an stronger tendency to valley
polarize, resulting in higher bands populating before the nearly-degenerate
bands.  Since the nearly degenerate $a$ bands are in the opposite valley from the
$a'$ bands, a higher $b$ band of the same valley may populate first, so that
the degenerate bands start filling at $\nu=1$ or $2$ as we see in panel (b)
rather than at $\nu=0$ in panels (c,d).  When this happens, the weight from
that higher band usually transitions downward as lower bands fill, causing the
diagonal lines we observe in all panels.  This can also decrease the range of
degenerate band filling from $4$ to $3$ or fewer, as some of these bands fill
by first-order transition instead of incrementally, which we see in panel (b).
Notably, the degeneracy of the blue spin in panel (b) is almost completely
obscured by this effect.  We also find in panel (c) that the 2/3 line can
be broken by jumps between the degenerate and higher bands, indicating that 
the state may be fragile.  From just the noninteracting picture, it is unintutive
that the higher band would be energetically close at this interaction strength.
We also find the 2/3 state more robust at higher interaction strength when the
higher band first fills completely, so we emphasize that the higher band gap 
criterion of Sec.~\ref{noninteracting} should not be taken too seriously.

To simulate the $SU(3)_2$ state, we compare the energy-minimized state to a
state with equal weight in each of the nearly-degenerate bands (\ie the four
$a$-bands or the four $b$-bands), whenever these bands have a combined filling
of 2/3.  We then keep this state for the lowest energy differences at which
it appears, assuming that an additional correlated energy gain of the $SU(3)_2$
state may overcome this energy difference.  For interaction energy $V=60$ and
$50$ kelvin in panels c and d, we keep the state if it is within $2/5\;V$ of
the minimum energy, $24$ and $20$ kelvin respectively.  For $V=100$ kelvin in
panel b, we keep the state if it is within $1/3 \; V$ of the minimum energy,
$33$ kelvin.  The presence of the correlated state is apparent from a vertical
line at 2/3+integer filling.  We see the integer can change depending on the
filling order, even in the middle of the degenerate range as in panel (c).  The
transition of weight from a higher band into the degenerate bands as they fill
can also remove the possibility of 2/3 filling, as we see for the blue spin in
panel (b).  For the models we considered, the $SU(3)_2$ Fibonacci state may
appear at $\nu= \frac23$ and $4\frac23$, $\nu=\frac23$ and $5\frac23$, or
$\nu=2\frac23$, always around horizontal white lines across consecutive fillings
of several Landau levels of the same spin.

Though it may be unlikely to realize a substrate coupling as large as $.045$ eV
to bring the degeneracy to zero displacement field, for completeness we discuss
what may occur in this scenario.  Here, four sets of degenerate bands fill at
the same displacement field, rather than pairs at positive and negative $u$.
Filling of the degeneracy will start from the low bands totally empty at $\nu=-8$.
Particle-like $2/3$ states may occur at $-7 \frac13$, $-3 \frac13$, $\frac23$, and $4\frac23$.
Integer shifts like we see for the finite $u$ degeneracy are unlikely, since the
higher bands are maximally far from all the degenerate bands at $u=0$.
Considering that the negative filling fraction (\ie negative Hall conductance) carriers 
are hole-like, we may instead expect a hole-like state, the particle-hole conjugate
of the $2/3$ state, to form at $-4\frac23$ and $-\frac23$ instead.  Thus when the
degeneracies are brought close enough to zero displacement field, we may find 
$SU(3)_2$ states at $\nu=\pm 4\frac23$ and $\pm \frac23$.

\section{Experimental signatures of non-Abelian FQH states}

Experimentally, to discover the Fibonacci FQH state in tetralayer graphene, one
may first locate either the $\nu_n = 0$ insulating state or the $\nu_n = \pm 4$
integer quantum Hall (IQH) state, depending on how large the substrate coupling
is estimated to be. From there, by changing the electron density,
one can attempt to access $\nu = \pm \frac{2}{3}$ or $\nu_n = \pm 4\frac{2}{3}$
primary stable FQH state, where the negative filling fraction states are only
relavant if the substrate coupling is strong and the degeneracy is near $u=0$.
If these are not found, particularly for weaker substrate coupling, one may
look above $\nu=2$ and $\nu=5$ for a $2\frac23$ or $5\frac23$ state.  

For the following discussion, to be more general we will also consider a hole-like 
(particle-like) state at positive (negative) filling fraction $\pm 3\frac13$.  We discuss what
happens for a single spin assuming the degeneracy fills from $\nu=0$, so that 
the states occur at $\pm \frac23$ and $\pm 3\frac13$.  This behavior can be 
extrapolated to the other spin at its own filling fractions, or to integer-shifted $\nu$ from 
higher bands filling first.

A $\nu_n = \pm \frac{2}{3}$ or $\nu_n = \pm 3\frac{1}{3}$ FQH state may be
either Abelian or non-Abelian.  Its nature can be inferred by examining other
related FQH states. For example, if $\nu_n = \pm \frac{1}{3}$ and $\nu_n = \pm
3\frac{2}{3}$ do not appear as primary stable FQH states, this suggests that
Landau-level degeneracy plays an important role, and the observed $\nu_n = \pm
\frac{2}{3}$ or $\nu_n = \pm 3\frac{1}{3}$ state may be non-Abelian. A more
reliable distinction can be made by probing the filling fractions of the
daughter states of the $\nu_n = \pm \frac{2}{3}$ or $\nu_n = \pm 3\frac{1}{3}$
FQH state \cite{ZW240612068}.  Specifically: 
\begin{itemize}

\item    If the daughter states appear at at $\nu_n = \pm 1$ and $\nu_n = \pm
\frac{7}{11}$ (for $\nu = \pm \frac{2}{3}$ state), or $\nu_n = \pm 3$ and
$\nu_n = \pm 3\frac{4}{11}$  (for $\nu = \pm 3\frac{1}{3}$ state), this
indicates that the parent state is the non-Abelian $SU(3)_2$ Fibonacci state.

\item    In contrast, if the daughter states occur at $\nu_n = \pm 1$ and
$\nu_n = \pm \frac{3}{5}$ or at $\nu_n = \pm 3$ and $\nu_n = \pm 3\frac{2}{5}$,
the parent state is an Abelian state, analogous to the particle-hole conjugate
of the $\nu = \frac{1}{3}$ Laughlin state.

\end{itemize}

If the Landau levels for a single species have three- or five-fold degeneracy,
we then look for primary stable FQH states at $\nu_n=\pm \frac12$ and
$\nu_n=\pm \frac35$ respectively.  If the $\nu_n =\pm \frac13$ state does not
appear as primary stable FQH states, this may be a hint that $\nu_n=\pm
\frac12$ and $\nu_n=\pm \frac35$ states are $SU(2)_2$ non-Abelian states
$\chi_1\chi_2^2$ and  $SU(2)_3$ non-Abelian state $\chi_1\chi_3^2$.


\section{Conclusion}

We use a realistic band structure to compute the Landau level spectrum of
multilayer rhombohedral graphene. Our goal is to determine the conditions under
which such systems can effectively simulate degenerate Landau levels of a
single species. In this regime, partially filled degenerate Landau levels can
give rise to non-Abelian FQH states, including those supporting Fibonacci
anyons in the four- and five-layer graphene systems.

While the single-species condition is essential for realizing the $SU(m)_n$
non-Abelian FQH states as described in Eq.~\eqref{nabLL}, we expect that
partial satisfaction of this condition can still lead to the formation of other
interesting FQH states, including alternative non-Abelian phases, within
partially filled degenerate Landau levels.

Our calculations also reveal that a positive value of $\Delta_2$ plays a
important role in enabling rhombohedral graphene to host non-Abelian FQH
states.  The optimal value of $\Delta_2$ is found to be approximately 45meV.
For a positive $\Delta_2$ less than the optimal value, the non-Abelian FQH
states can still be realized, but with a reduced range of magnetic fields.
Therefore, it is important to both theoretically understand and experimentally
investigate how the substrate influences $\Delta_2$, and to explore methods to
tune or control this parameter in real devices.

\

\

We would like to thank Long Ju, Patrick Ledwith, Andrea Young, Yuan-Bo Zhang,
and Jun Zhu for very helpful discussions and comments.  X.-G.W was partially
supported by NSF grant DMR-2022428 and by the Simons Collaboration on
Ultra-Quantum Matter, which is a grant from the Simons Foundation (651446,
XGW).   AT was supported by NSF Graduate Research Fellowship grant number
2141064.

\bibliography{all,publst}

\appendix

\end{document}